%% file: recsys2024.tex
\documentclass[sigconf,natbib=true]{acmart}

\usepackage{subfigure}
\usepackage{float}
\usepackage{multirow}
\usepackage{enumitem}
\usepackage{setspace}
\usepackage{xspace}
\usepackage{bm}

\newcommand{\paratitle}[1]{\smallskip\noindent\textbf{#1}}
\newcommand{\ie}{\emph{i.e.,}\xspace}
\newcommand{\eg}{\emph{e.g.,}\xspace}

\AtBeginDocument{%
  }

\copyrightyear{2024}
\acmYear{2024}
\setcopyright{acmlicensed}
\acmConference[RecSys '24]{18th ACM Conference on Recommender Systems}{October 14--18, 2024}{Bari, Italy}
\acmBooktitle{18th ACM Conference on Recommender Systems (RecSys '24), October 14--18, 2024, Bari, Italy}
\acmDOI{10.1145/3640457.3688113}
\acmISBN{979-8-4007-0505-2/24/10}

\begin{document}
\title{Not All Videos Become Outdated: Short-Video Recommendation by Learning to Deconfound Release Interval Bias}

\author{Lulu Dong}
\affiliation{
  \institution{Faculty of Economics and Management, East China Normal University}
  \city{Shanghai}
  \country{China}
}
\email{lldong@stu.ecnu.edu.cn}

\author{Guoxiu He}
\authornote{Corresponding author.}
\affiliation{
  \institution{Faculty of Economics and Management, East China Normal University}
  \city{Shanghai}
  \country{China}
}
\email{gxhe@fem.ecnu.edu.cn}

\author{Aixin Sun}
\affiliation{
  \institution{College of Computing and Data Science, Nanyang Technological University}
  \city{Singapore}
  \country{Singapore}
}
\email{axsun@ntu.edu.sg}

\renewcommand{\shortauthors}{Lulu Dong et al.}

\begin{abstract}

Short-video recommender systems often exhibit a biased preference to \textit{recently released} videos.
However, not all videos become outdated; certain classic videos can still attract user's attention. 
Such bias along temporal dimension can be 
further aggravated by the matching model 
between users and videos, because the model learns from preexisting interactions.
From real data, we observe that different videos have varying sensitivities to recency in attracting users' attention. 
Our analysis, based on a causal graph modeling short-video recommendation, suggests that the \textit{release interval} serves as a confounder, establishing a backdoor path between users and videos.
To address this confounding effect, we propose a model-agnostic causal architecture called \textbf{L}earning to \textbf{D}econfound the \textbf{R}elease \textbf{I}nterval Bias (\textbf{LDRI}). LDRI enables jointly learning of the matching model and the video recency sensitivity perceptron. In the inference stage, we apply a backdoor adjustment, effectively blocking the backdoor path by intervening on each video. Extensive experiments on two benchmarks demonstrate that LDRI consistently outperforms backbone models and exhibits superior performance against state-of-the-art models. Additional comprehensive analyses confirm the deconfounding capability of LDRI.
\end{abstract}

\ccsdesc[500]{Information systems~Recommender systems}

\keywords{Recommender System, Causal Inference, Release Interval Bias} 

\maketitle
\input{contents/01intro}
\input{contents/02relatedwork}
\input{contents/03causal}
\input{contents/04method}

\input{contents/05experiments}
\input{contents/06results}
\input{contents/07conclusion}

\begin{acks}
This work is supported by the National Natural Science Foundation of China (72204087), the Shanghai Planning Office of Philosophy and Social Science Youth Project (2022ETQ001), the "Chen Guang" project supported by Shanghai Municipal Education Commission and Shanghai Education Development Foundation (23CGA28), the Shanghai Pujiang Program (23PJC030), the Fundamental Research Funds for the Central Universities, China, and the 2024 Innovation Evaluation Open Fund, Fudan University (CXPJ2024006). We also appreciate the constructive comments from the anonymous reviewers.
\end{acks}

\bibliographystyle{ACM-Reference-Format}
\bibliography{recsys2024}

\end{document}

%% file: contents/01intro.tex
\section{Introduction}
\label{sec:intro}

\begin{figure}[t]
  \centering
  \subfigure[Exposure distribution]{
    \centering
    \includegraphics[width=0.425\columnwidth]{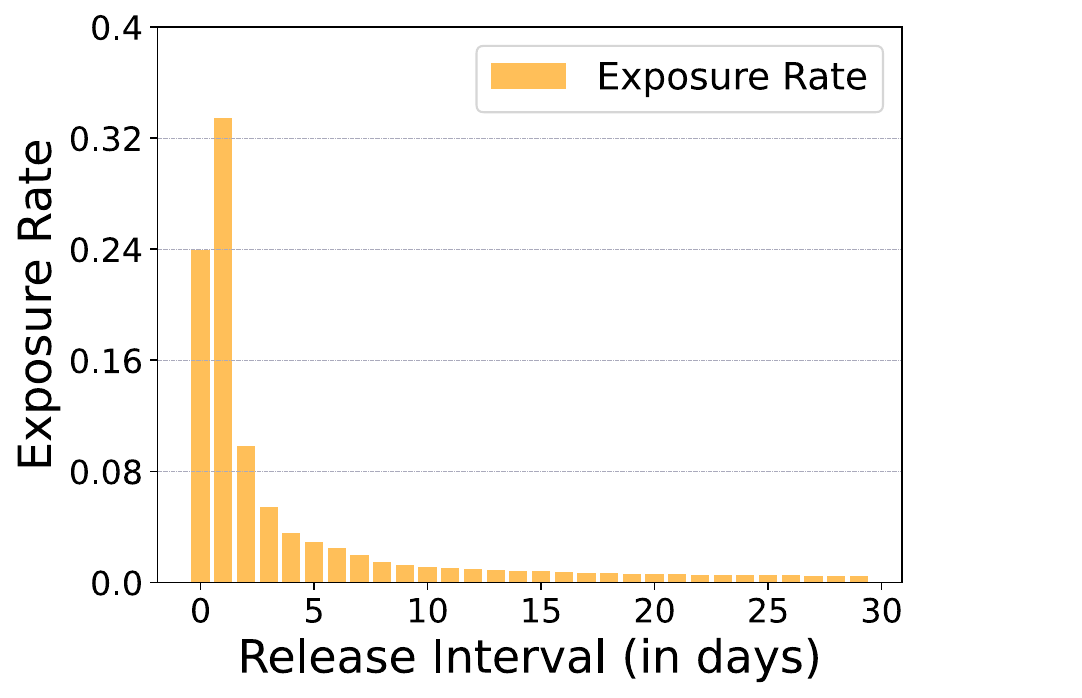} 
    \label{fig:expose}}
  \quad
  \subfigure[Positive feedback distribution]{
    \includegraphics[width=0.425\columnwidth]{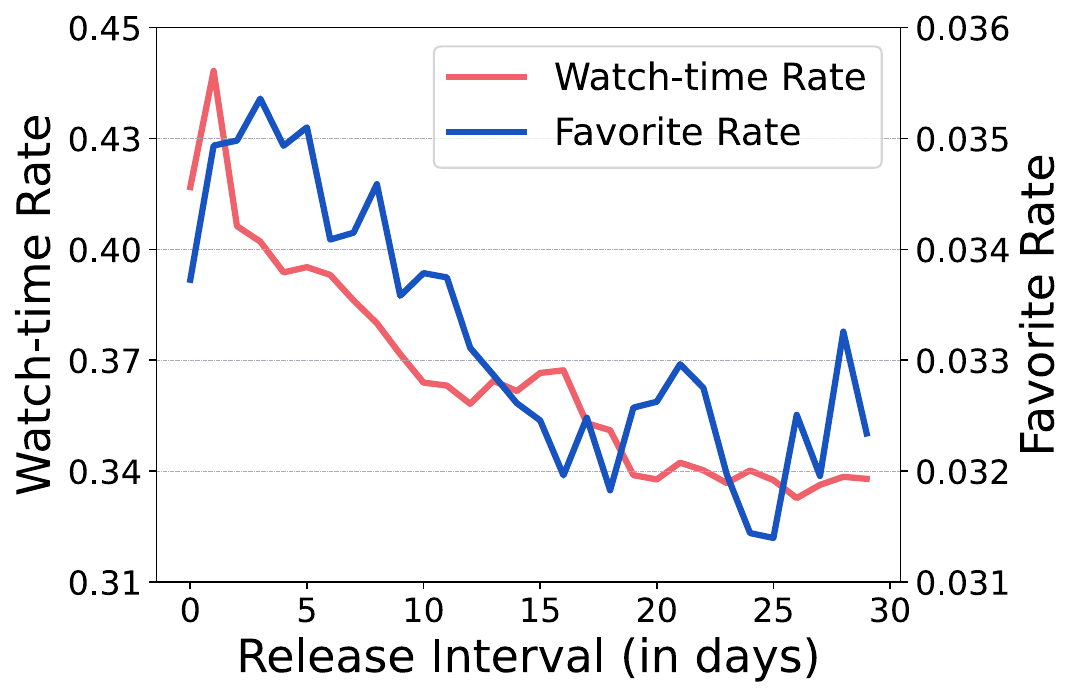}
    \label{fig:interact}}
    \quad
    \subfigure[Topic 2: recency-insensitive]{
  \includegraphics[width=0.425\columnwidth]{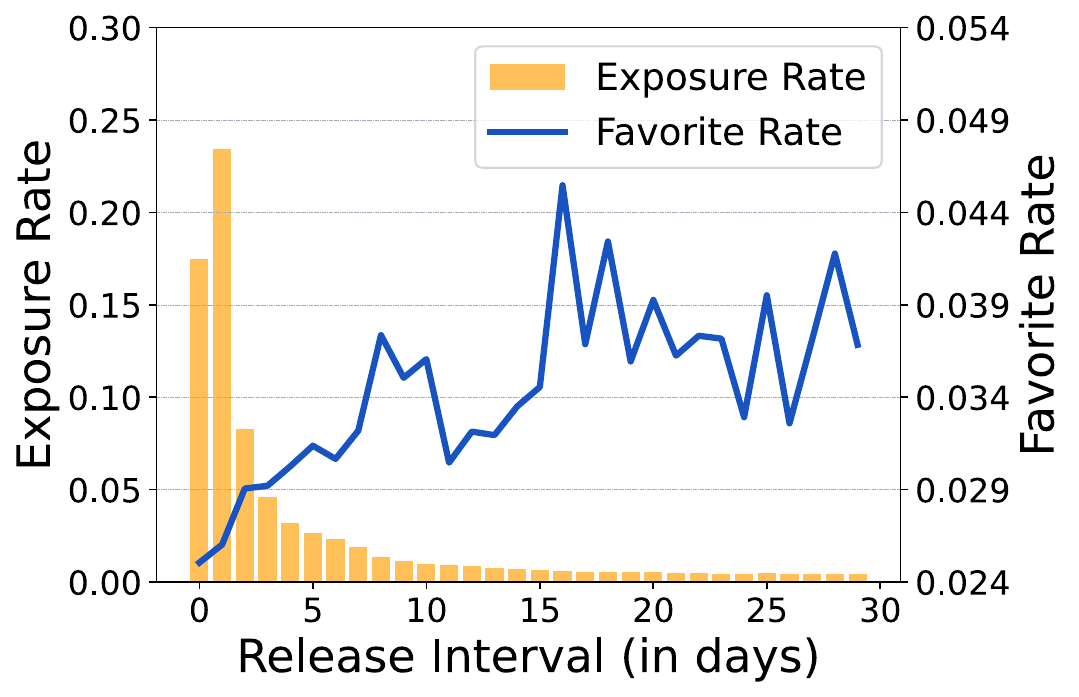}
  \label{fig:tag2}}
  \quad
\subfigure[Topic 20: recency-sensitive]{
  \centering
  \includegraphics[width=0.425\columnwidth]{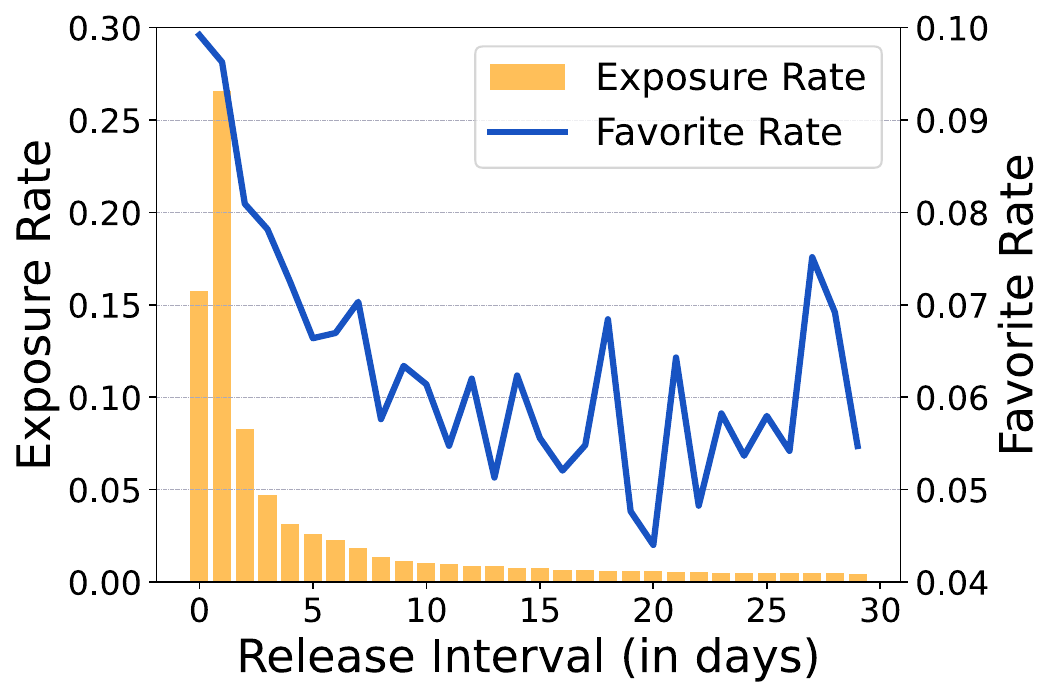}
  \label{fig:tag20}}
    
  \caption{(a) and (b) plot video exposure rate and user positive feedbacks against video release interval in days, on all records of KuaiRand-1K. (c) and (d) show the trend of user positive feedback for videos on two different topics with similar exposure distribution. Watch-time rate is the ratio of the watch time to the video length; favorite rate represents the ratio of users who liked a video to those who watched it.
  }
  \label{fig:rate}
\end{figure}

The emergence of user-generated short-video platforms such as Youtube Shorts and Kuaishou has been catalyzing the prosperity of the self-media landscape~\cite{shutsko2020user}. 
Short-video recommender systems leverage user-video interaction data to develop sophisticated user interest models~\cite{cai2023reinforcing,tian2022multi,covington2016deep,lin2022feature}, aiming to accurately estimate matching scores between users and videos. 

Nevertheless, current short-video recommender systems tend to exhibit bias towards \textit{recently released} videos.
We analyze a real-world dataset called KuaiRand-1K, collected from the logs of the Kuaishou app~\cite{gao2022kuairand}.
Figure~\ref{fig:expose} depicts the decrease in video exposure as the video release interval increases. 
Figure~\ref{fig:interact} demonstrates a decline in positive user feedback with videos, \eg watch-time and favorite rate,  along the release interval increases. It is important to note that both figures are plotted based on all records in the dataset. 
With the platform's aim of maximizing user positive feedback, models training on such biased interactions can potentially capture a shortcut between short release intervals and positive user feedback, thereby further exacerbating the existing bias \cite{mansoury2020feedback,chaney2018algorithmic}.

However, not all videos become outdated over time. Videos that are less sensitive to recency, \eg fitness videos, maintain their relevance over time. 
Although there is no semantic labels of the topics, videos in the KuaiRand-1k dataset are tagged with topic IDs. Figure~\ref{fig:tag2} (Topic ID 2) displays an upward trend in positive user feedback, implying that videos on this topic are less sensitive to recency. Recommending these videos, even if they have been released for an extended period, will contribute positively to user experience. As a comparison, Figure~\ref{fig:tag20} (Topic ID 20) shows a typical decline in user interest in videos over time, suggesting that videos on this topic are sensitive to recency. 
However, the recommender employs the same exposure mechanism for both topics, prioritizing videos with a shorter release interval. 
Addressing the bias towards newly released videos and providing tailored recommendation based on unique recency sensitivity is urgently needed.

Existing efforts~\cite{chen2023bias} to mitigate biases in short-video recommender systems have shown progress in addressing biases like popularity~\cite{tang2023counterfactual,zhang2021causal} and duration~\cite{zhan2022deconfounding}. 
Some recent work have focused on the modeling of recency sensitivity characteristics of videos~\cite{chen2023tccm,10.1145/3488560.3498375,zhang2017building}; nonetheless, they have not recognized the bias stemming from the release interval of videos.
Survival analysis emerges as the principal methodology in these endeavors, which allows for a comprehensive understanding of the temporal dynamics of videos involved in user-item interactions~\cite{huang2023personal,wang2023measuring}. 
Unfortunately, mainstream methods in this domain, such as the Cox model~\cite{therneau2000cox}, often rely on manually crafted rules to determine video deactivation and typically exclude videos that have not been recorded as deactivated during training and testing~\cite{wang2023measuring}. This renders them impractical in real-world recommendations, necessitating further modeling of recency sensitivity. 

We name the bias towards videos with short release intervals as \textit{release interval bias}; to address this bias, 
we resort to causal inference~\cite{pearl2009causality}. We first formalize the current recommendation process as a causal graph (detailed in Section~\ref{sec:3}). Through this analysis, we identify the shortcut between videos and the predicted user interests, which is a non-causal relationship confounded by \textit{release interval}. 
To eliminate the confounding effect, we propose a model-agnostic framework called \textbf{L}earning to \textbf{D}econfound \textbf{R}elease \textbf{I}nterval Bias (\textbf{LDRI}).  

In training stage, LDRI enables joint-learning of video recency representation and matching model that connects user features with video contents. To accomplish this, we design a recency perceptron that captures each video's recency sensitivity at every release interval by leveraging the inherent features of videos and users' feedback. The streamlined yet effective design enables its application to any matching model without additional feature engineering and data filtering. During inference, LDRI succeeds in mitigating confounding effect through backdoor adjustment.
Extensive experiments on two benchmarks from real-world scenarios reveal that LDRI consistently outperforms three well-established backbone models, namely, DeepFM~\cite{guo2017deepfm}, NFM~\cite{he2017neural}, and AFM~\cite{xiao2017attentional}. In addition, LDRI achieves superior performance compared with state-of-the-art models considering the duration or temporality factor.
Comprehensive analyses provide valuable insights into LDRI's deconfounding ability.
Our contributions are summarized as follows:
\begin{itemize}[nolistsep, leftmargin=*]
    \item We approach user interests for videos from two perspectives namely the user-video matching and video recency. A model-agnostic architecture is designed to jointly learn both aspects.
    \item We demonstrate that release interval is a confounder in short-video recommender systems, and propose a causal framework to eliminate the confounding effect through backdoor adjustment.
    \item Extensive experiments on two large datasets have demonstrated the prominence and generality of our solutions, and the deconfounding capability.
\end{itemize}

%% file: contents/02relatedwork.tex
\begin{figure*}[]
  \centering
  \subfigure[Traditional process]{
    \centering
    \includegraphics[width=0.3\columnwidth]{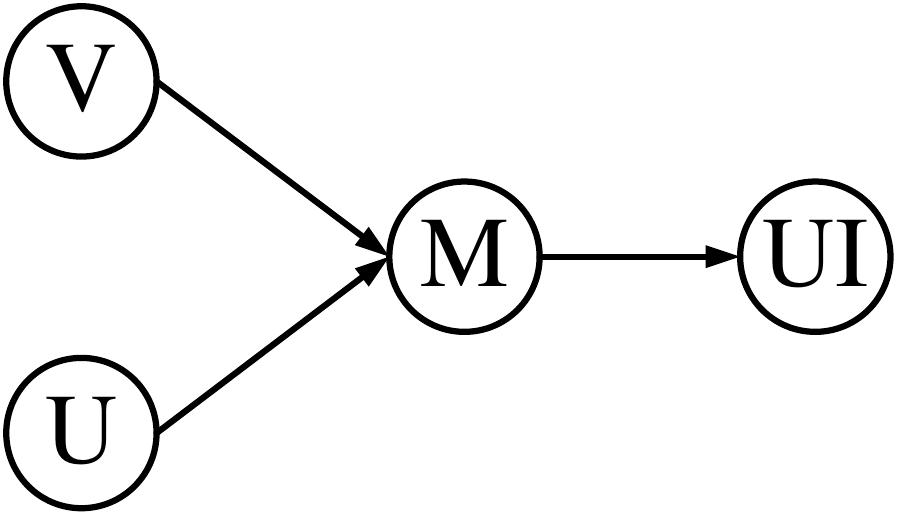}
    \label{fig:causal1}}
  \quad
  \quad
  \quad
  \subfigure[Without confounder]{
    \includegraphics[width=0.3\columnwidth]{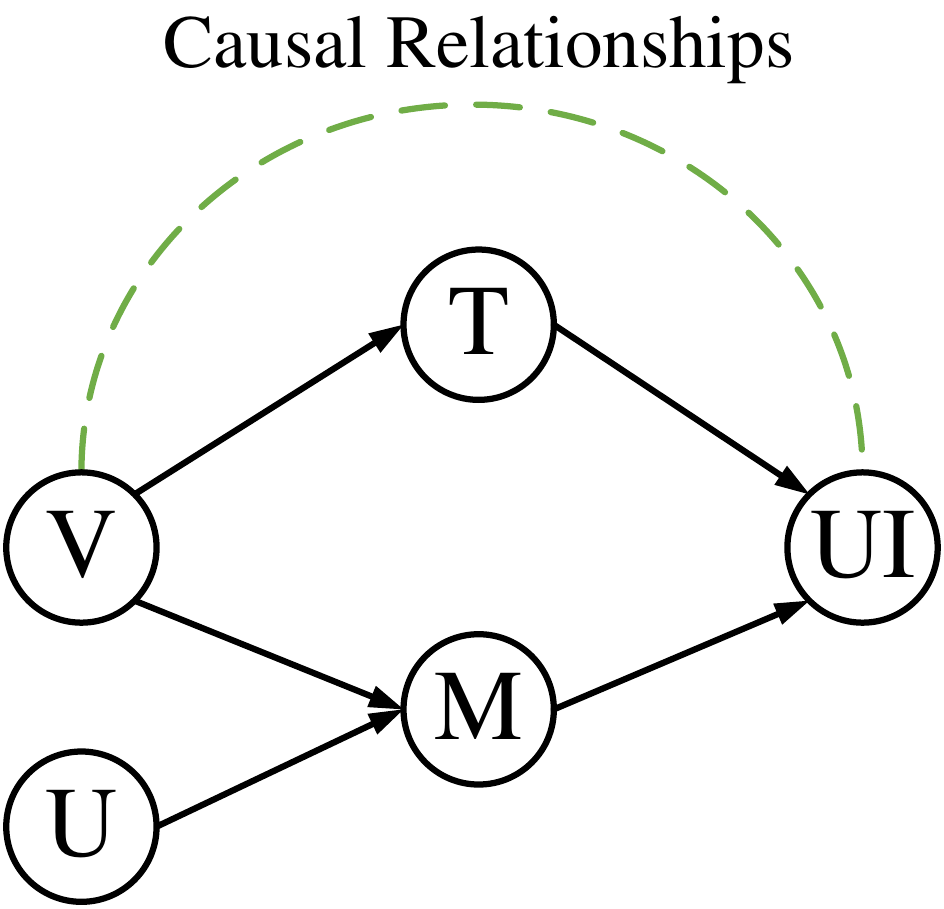}
    \label{fig:causal2}}
  \quad
  \quad
  \quad
  \subfigure[With confounder]{
    \includegraphics[width=0.3\columnwidth]{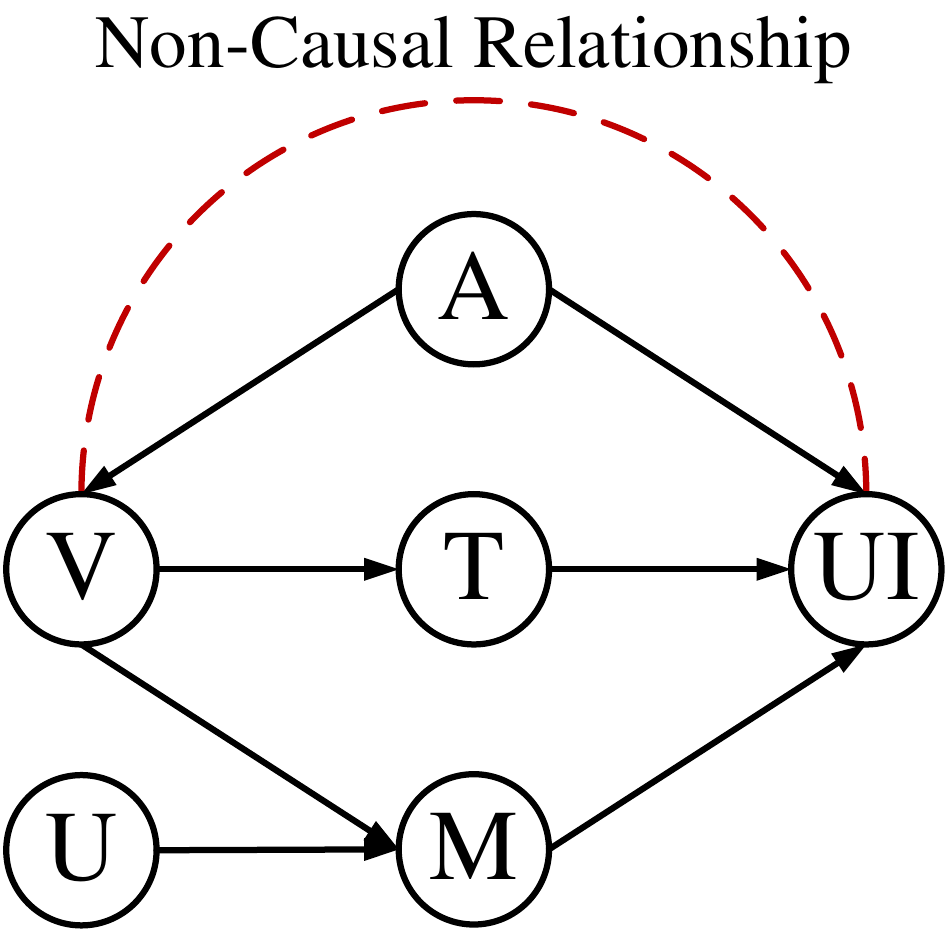}
    \label{fig:causal3}}
  \quad
  \quad
  \quad
  \subfigure[Backdoor adjustment]{
  \includegraphics[width=0.3\columnwidth]{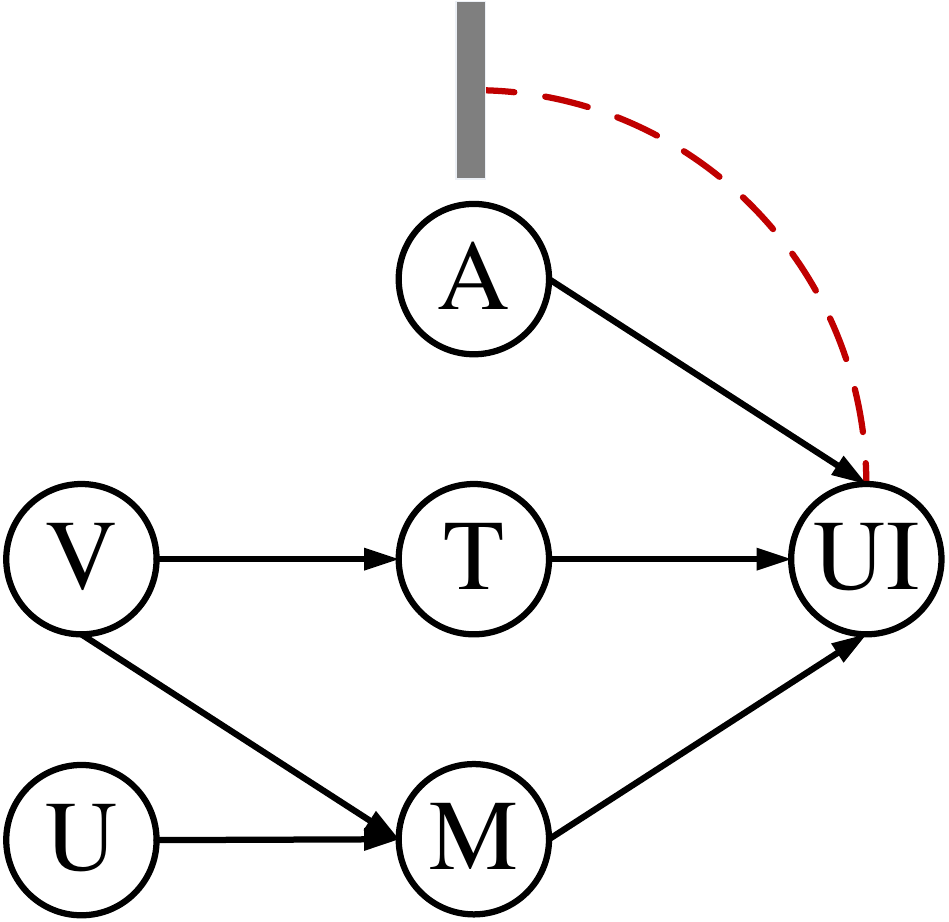}
    \label{fig:causal4}}
  \caption{Causal view of short-video recommendations. These causal graphs represent traditional recommendation, ideal recommendation incorporating recency sensitivity of video, recommendation confounded by release interval bias, and decoufounded recommendation through backdoor adjustment, respectively. Nodes in the graphs are: video features ($V$), user features ($U$), matching between video and user ($M$), user interests ($UI$), recency sensitivity of video ($T$), and release interval confounder ($A$).}
  \Description{illustration of the factors.}
  \label{fig:causal}
\end{figure*}
\section{Related Work}
\label{sec:relate}

\subsection{Bias in Recommendation}
Recommender systems often suffer from various biases that can significantly impact both the performance of the recommendations and user experience.
These biases primarily arise from data non-i.i.d., model design, and feedback-loop nature of recommendation scenarios \cite{mansoury2020feedback,chen2023bias,wang2021deconfounded}. The o.o.d. problem is a major contributor to selection bias, conformity bias, sentiment bias, \textit{etc} \cite{jeunen2023probabilistic,he2022causal,lin2021mitigating}. In certain scenarios, biases may be desirable, leading some studies to intentionally introduce them into models~\cite{ding2019reinforced}. However, the feedback-loop nature of recommendation can further intensify biases~\cite{mansoury2020feedback,chaney2018algorithmic}, as the training of recommender systems heavily relies on existing interactions. Thus, models possess the potential to capture the shortcuts from biased data, further amplifying biases~\cite{tang2023counterfactual}.
A unique bias in short-video recommender systems, known as duration bias, has attracted significant attention in recent research \cite{he2023addressing,tang2023counterfactual,zhan2022deconfounding,quan2023alleviating,zheng2022dvr}. This bias misguides the recommender towards favoring either short- or long-duration videos exclusively. Another focal point in recent studies is popularity bias, which can lead to a rich-get-richer scenario~\cite{gao2023alleviating,zhang2021causal,wei2021model,shi2023relieving}. 

To the best of our knowledge, little prior work has tackled the shortcut caused by release interval bias in short-video recommender systems. 
To address this bias, we identify the \textit{release interval confounder} as the underlying source and resort to causal inference to mitigate its confounding effect.

\subsection{Item Temporality Modeling}

The temporality of items has historically been overlooked, but recently gained attention as a pivotal attribute. \citet{chen2023tccm} incorporated the release interval of items as a feature in their models for training and prediction. 
\citet{zhang2017building} quantified item's timeliness by measuring its similarity to the most popular items as its timeliness embedding. The age of items has been introduced to assess their dynamic effect on selection bias and user preference~\cite{10.1145/3488560.3498375}.
In addition to modeling from intrinsic features, survival analysis examines the time until the occurrence of an interactive event~\cite{jenkins2005survival,wang2019machine}.
It estimates item's probability of keeping active until events occur as its global temporality information~\cite{huang2023personal,wang2023measuring}. Nonetheless, this approach is limited to strictly filtered data and overlooks the internal item information.

In contrast to existing studies, we propose a model-agnostic architecture that leverages both user feedback and item features to learn the recency sensitivity of each item.

\subsection{Causal Recommendation}

Causal inference has emerged as a central focus in recent research on recommender systems~\cite{chen2023bias,pearl2009causality}. Ongoing studies in this field have demonstrated its prominent performance in addressing biases, as nearly all taxonomies of biases can be explained through a causal lens~\cite{chen2023bias}. Mainstream causal debiasing methods based on causal graph \cite{pearl2009causality} (acyclic directed graph consisting of nodes and edges) can be categorized into following types: 
(i) Counterfactual Interference \cite{tang2023counterfactual,wei2021model}. This method decomposes total effect (TE) of target nodes on outcome into natural direct effect (NDE) and total indirect effect (TIE). 
(ii) Backdoor Adjustment \cite{he2023addressing,zhan2022deconfounding,lin2023tree}. 
An essential part of implementing backdoor adjustment involves recognizing confounding variables. These variables have the potential to create non-causal connections between treatments and outcomes, which can exacerbate unforeseen biases.
\citet{zhang2021causal} regards item popularity as a confounder between items and users' feedback. By applying backdoor adjustment, the amplification of implicit biases is prevented.

In this paper, we employ backdoor adjustment to eliminate the non-causal relationship introduced by \textit{release interval} confounder in the recommender system.

%% file: contents/03causal.tex
\section{Deconfounding Recommendation} 
\label{sec:3}
In this section, we analyze short-video recommendations from a causal perspective. Then, we present our causal framework \textbf{LDRI}. 

\subsection{Causal View of Recommendation}
\label{sec:3.1}
Causal inference focuses on analysing both causal and non-causal relationships between variables within a system. Non-causal relationships often arise when confounding factors are present, leading to a high correlation between variables that are not causally related. These relationships complicate the identification of true causal relationships and pose challenges for establishing causality. One prominent tool for analysing such intricate relationships is the causal graph. Causal graph is a directed acyclic graph (DAG), where nodes represent variables, and directed edges depict the causal relationships between them.

Figure~\ref{fig:causal} illustrates 4 causal graphs describing different recommendation processes. We elaborate edges in each graph as follows:
\begin{description}
  \item[$\{U,V\} \rightarrow M \rightarrow UI$:] the matching between user features and video contents entails that the video could captivate users' attention, leading to a causal effect of fulfilling their interests.
  \item[$V \rightarrow T \rightarrow UI$:] video possesses an innate sensitivity to recency, resulting in a direct causal effect on user interests, as outdated video content fails to captivate users.
  \item[ $V \rightarrow T \rightarrow UI \leftarrow M \leftarrow \{U, V\}$:] the total causal effects of the video on user interests is decomposed into the above two paths.
  \item [$V\leftarrow A \rightarrow UI$:] the backdoor path behind $V$ and $UI$ opened by confounder $A$, \ie \textit{release interval}. This confounder introduces non-causal relationship between video and user interests, as illustrated by the dashed red line in this graph.
\end{description}

Particularly, Figure \ref{fig:causal1} presents the traditional simplified recommendation that estimates the matching score between user features and video contents. However, this approach ignores the inherent recency sensitivity of videos. Outdated videos may not align with users' interests. 
Figure~\ref{fig:causal2} illustrates an ideal recommendation which incorporates the recency sensitivity of videos. Here, the total causal effects of videos on user interests emanates from two paths; videos must (i) capture users' attention, and (ii) remain relevant by not being outdated.
However, as discussed in Section~\ref{sec:intro}, realistic recommendations are often exposed to the bias of preferring to recommend recently released videos, regardless of whether other videos are actually outdated. Figure~\ref{fig:causal3} recognizes a confounding factor, the \textit{release interval} confounder. This previously overlooked confounder creates a non-causal shortcut between videos with short release intervals and user positive feedback, thereby confounding the recommendation process.

In this work, we aim to address the confounding effect introduced by the \textit{release interval} confounder. To this end, we employ backdoor adjustment~\cite{pearl2009causality} to block the backdoor path opened by confounder, as shown in Figure~\ref{fig:causal4}, which eliminates non-causal relationships and only retains the causal relationships.
Section~\ref{sec:3.2} gives the derivation of backdoor adjustment and elaborates our causal model.

\subsection{LDRI: Deconfounding by Backdoor Adjustment}
\label{sec:3.2}
Conceptually, the causal effects of $V$ on $UI$ is the difference of $UI$ when the value of $V$ changes from reference status $V=v^*$ to target status $V=v$, namely intervening on $V$. In the preliminaries of \textit{do-calculus}, we formulate the prediction model as follows:
\begin{equation}
\label{eqn3.1}
  P(UI|do(V=v)) - P(UI|do(V=v^*))
\end{equation}
where $do(V)$ means intervening on variable $V$. By intervening on $V$, the edge from $A \to V$ can be removed. Accordingly, the backdoor path is blocked and non-causal relationship is eliminated from causal graph $G$, as shown in Figure~\ref{fig:causal4}. $P(UI|do(V=v^*))$ is often treated as zero or constant value that can be ignored since recommendation task is a ranking task. Hence, modeling Equation~(\ref{eqn3.1}) is equal to modeling: 
\begin{equation}
\label{eqn3.2}
  P\left(UI|do(V=v)\right)
\end{equation}

We frame the causal deconfounding prediction model as Equation (\ref{eqn3.2}) and derive it from the joint distribution of $\{UI, A, M, T, V\}$ in graph $G$:
\begin{equation}
\label{eqn3.3}
  P(UI, A, M, T, V)\overset{\text{(i)}}{=}P(UI|A, M, T)P(M|U, V)P(T|V)P(V|A)P(A)
\end{equation}
By intervening on $V$, we have:
\begin{equation}
\label{eqn3.4}
    P(UI, A, M, T|do(V))\overset{\text{(ii)}}{=}P(UI|A, M, T)P(M|U, V)P(T|V)P(A)
\end{equation}
Furthermore:
\begin{equation}
\label{eqn3.5}
  P(UI|do(V))\overset{\text{(iii)}}{=}\sum_A\sum_M\sum_TP(UI|A, M, T)P(M|U, V)P(T|V)P(A)
\end{equation}

Equations (\ref{eqn3.3}) - (\ref{eqn3.5}) demonstrate the derivation of Equation (\ref{eqn3.2}) that implements backdoor adjustment. In these equations, 
(i) is by the assumption of causal inference that one node depends only on its parent nodes;
(ii) is because $P(V|A)=1$ when intervene on $V$; 
(iii) is by the law of marginal distribution.
By employing backdoor adjustment, the edge from $A \to V$ is removed, thereby blocking the backdoor path $V\leftarrow A \rightarrow UI$. Consequently, the non-causal relationship between $V$ and $UI$ is eradicated, leaving only causal relationships. Drawing on Equation~(\ref{eqn3.5}), recommendation can be formulated based on two videos' causal effects on user interests.

In the context of short-video recommendation, we examine a specific interaction log denoted as $(u, v, a, y)$, where $u$ represents user features, $v$ represents video contents, $a$ denotes release interval, and $y$ represents user feedback. We employ a matching model between $u$ and $v$ to estimate matching score $m$ for which a video attracts user's attention. Furthermore, we incorporate a perceptron to discern the recency sensitivity of $v$. The sensitivity variations of a video across release intervals are encapsulated in a vector $\bm{t}$, where each dimension of $\bm{t}$ delineates its recency sensitivity at corresponding release interval. 
Therefore, Equation (\ref{eqn3.5}) can be formulated as follows:
\begin{equation}
\label{eqn3.6}
  P(UI|do(v))=\sum_AP(UI|A, m, \bm{t})P(A)
\end{equation}
During the inference stage, we design a fusion function to employ $m$ and $\bm{t}$ to make recommendation, corresponding to the two paths of causal effect from $V$ to $UI$: $V \rightarrow T \rightarrow UI$ and $\{U,V\} \rightarrow M \rightarrow UI$. 
We formulate the recommendation as follows: 
\begin{equation}
\label{eqn3.7}
  P(UI|do(v))=\sum_A\mathcal{F}\left[A, m=\mathcal{M}(u, v), \bm{t}=\mathcal{T}(v)\right]P(A)
\end{equation}
where $\mathcal{M}(\cdot), \mathcal{T}(\cdot), \mathcal{F}(\cdot)$ denote a matching model between user features and video features to estimate relevance score $m$, a perceptron of video recency sensitivity to extract representation $\bm{t}$, and a fusion function to conduct final recommendation, respectively. 

Particularly, we implement two inference policies that are widely adopted by existing studies:
 
 \paratitle{Inference Policy 1.} In line with work \cite{zhang2021causal,zhan2022deconfounding}, we can formulate Equation (\ref{eqn3.7}) as follows:
  \begin{equation}
    \label{eqn3.8}
    \begin{aligned}
      P(UI|do(v)) &\approx \sum_a \textbf{1}\{A=a\} \mathcal{F}\left[A=a, m=\mathcal{M}(u, v), \bm{t}=\mathcal{T}(v) \right] \\
    \end{aligned}
    \end{equation}

  \paratitle{Inference Policy 2.} In line with work \cite{he2023addressing}, this inference policy iterates each value of $A$ to make inference:
  \begin{equation}
    \label{eqn3.9}
      P(UI|do(v))=\sum_a\mathcal{F}\left[A=a, m=\mathcal{M}(u, v), \bm{t}=\mathcal{T}(v)\right]P(A=a)
    \end{equation}
    where $P(A=a)$ can be estimated by the frequency of $A=a$ in dataset $\mathcal{D}$, treated as a global value.

We denote the inference derived from \textit{Policy 1} as \textit{LDRI}, while the the inference from \textit{Policy 2} is indicated as \textit{LDRI-iter}. We will introduce the detailed instantiation of our proposed causal recommendation framework in Section~\ref{sec:4}. 

%% file: contents/04method.tex
\section{Instantiation}
\label{sec:4}

\subsection{Preliminaries}
\label{sec:4.1}

\begin{figure*}[t]
    \centering
    \includegraphics[width=0.85\textwidth]{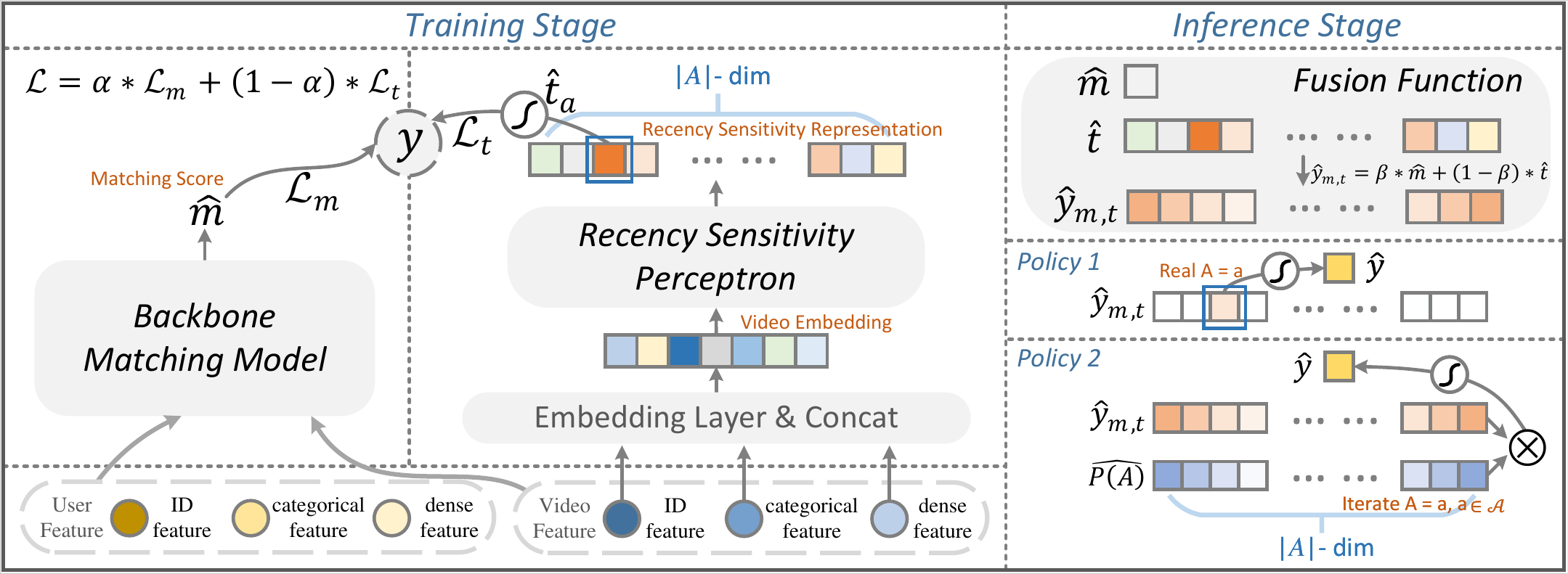}
    \caption{Overview of the proposed LDRI framework. The left and middle parts illustrate the joint training of matching model between user features and video features, as well as perceptron of recency sensitivity in training stage. The right part depicts the inference stage, where LDRI fuses user interests and recency sensitivity of video, and implements backdoor adjustment to mitigate the confounding effect introduced by release interval.}
    \label{fig:model}
\end{figure*}

We denote user feature set as $\mathcal{U}$, video feature set as $\mathcal{V}$, and release intervals as $A=\{0, 1, ..., \lvert A \rvert -1 \}$, which is quantified in days. We employ binary training label, denoted as $y$, to indicate user feedback. The historical interaction logs in our dataset are represented as $\mathcal{D}=\left\{(u, v, a, y)|u \in \mathcal{U},v \in \mathcal{V},a \in \mathcal{A}, y \in \{0, 1\}\right\}$. During the training stage, as described in Equation (\ref{eqn3.7}), our goal is to estimate a matching score $m$ for each pair of user $u$ and video $v$, using the matching model $\mathcal{M}(u, v)$. Additionally, we aim to extract a recency sensitivity vector $\bm{t}$ for each video $v$ using the recency perceptron $\mathcal{T}(v)$. During the inference stage, we design a fusion function $\mathcal{F}(A, \bm{t}, m)$ and employ two inference policies to make final recommendation. Our proposed LDRI is depicted in Figure \ref{fig:model}.

\subsection{Training Stage}
\label{sec:4.2}
During the training stage, we propose a multi-task learning framework that jointly trains the models: $\mathcal{M}(u, v)$ and $\mathcal{T}(v)$.

\subsubsection{Matching model} 
$\mathcal{M}(u, v)$ can be any recommendation model to predict whether the video can attract users' attention. As demonstrated in the left part of Figure~\ref{fig:model}, $\mathcal{M}(u, v)$ is a backbone of LDRI to estimate the matching score $m$. For an interaction log $(u, v, a, y)\in \mathcal{D} $, we embed user features as $\bm{u}\in\mathbb{R}^{d_u}$ and video features as $\bm{v}\in\mathbb{R}^{d_v}$ and input them into $\mathcal{M}$ to conduct prediction. The predicted score derived from $\mathcal{M}$ is denoted as $\hat{m}$. 
Since this is a binary classification task, the loss can be calculated by binary cross-entropy (BCE):
\begin{equation}
  \label{eqn4.1}
  \mathcal{L}_m = BCE_m(\hat{m}, y)
\end{equation}

\subsubsection{Perceptron of recency sensitivity}
\label{sec:4.2.2}

The middle part of Figure~\ref{fig:model} depicts our perceptron of recency sensitivity $\mathcal{T}(v)$, which is devised to capture each video's recency sensitivity. 
Since \textit{V} $\rightarrow$ \textit{T} in Figure \ref{fig:causal3} implies that a video possesses its inherent recency sensitivity, the intuitive approach is to derive the recency sensitivity vector $\bm{t}$ from its own features. In this light, we embed video feature as $\bm{v} \in \mathbb{R}^d $ and project it into a $\lvert A \rvert$-dimensional vector:
\begin{equation}
  \label{eqn4.2}
  \begin{aligned}
    \bm{\hat{t}} = \mathcal{T}(\bm{v}), 
    \bm{\hat{t}} \in \mathbb{R}^{\lvert A \rvert}
  \end{aligned}
\end{equation}
Here, we employ the Res-MLP \cite{he2016deep} as the projector. We denote the $a_{th}$ dimensional $\bm{\hat{t}}_a$ to present the video's recency sensitivity at $A=a$. 
Given that the recency sensitivity of a specific video at adjacent release intervals may exhibit correspondence, $\bm{\hat{t}}_a$ is supervised by user positive feedbacks of nearby neighbors.
Thus, we calculate the BCE loss as follows:
\begin{equation}
\label{eqn4.4}
\begin{split}
    &\mathcal{L}_t = BCE_t(\sigma(\bm{\hat{t}}_a^{\prime}), y)
    \\
    &\bm{\hat{t}}_a^{\prime} = \frac{\bm{\hat{t}}_{a-N}+ \cdots + \bm{\hat{t}}_{a} + \cdots + \bm{\hat{t}}_{a+N}}{2\ast N+1}
\end{split}
\end{equation}
where $2\ast N+1$ denotes the window size and $\sigma(\cdot)$ denotes \textit{sigmoid} function. $y\in \{0, 1\}$ indicates the training label that whether users give positive feedback at $A=a$. Thus, $\bm{\hat{t}}_a$ and its adjacent dimensions can be supervised by user positive feedback together. In this way, the larger value of $\bm{\hat{t}}_a$ means that the video receives more positive feedback around $A=a$, suggesting a lower probability of the video becoming outdated. Despite having limited feedback or recently released videos, the proposed perceptron can still be effectively trained using the complete set of interaction data. This enables the perceptron to accurately estimate recency sensitivity at each release time value.

\subsubsection{Joint Learning}
LDRI employs a joint learning approach to train the two models.
Hence, the training loss is formulated as:
\begin{equation}
  \label{eqn4.6}
  \begin{aligned}
    \mathcal{L} &= \alpha \times \mathcal{L}_m + (1-\alpha) \times \mathcal{L}_t \\
                &= \alpha \times BCE_m(\hat{m}, y) + (1-\alpha) \times BCE_t(\sigma(\bm{\hat{t}}_a), y)
  \end{aligned}
\end{equation}
where $\alpha \in (0, 1)$ is the hyper-parameter to balance the training of two models. The left and middle parts of Figure~\ref{fig:model} represent the joint learning approach of LDRI. 

\subsection{Inference Stage}
\label{sec:4.3}
As illustrated in the right part of Figure~\ref{fig:model}, during the inference stage, we fuse the estimated matching score $\hat{m}$ and extracted recency sensitivity representation $\bm{\hat{t}}$ to conduct inference. The fusion function $\mathcal{T}(A, \hat{m}, \bm{\hat{t}})$ is formulated as follows:
\begin{equation}
  \label{eqn4.7}
  \hat{\textbf{y}}_{m, t} = \beta \times \hat{m} + (1-\beta) \times \bm{\hat{t}}
\end{equation}
where $\beta \in (0, 1)$ is the weight balancing user-video matching and recency sensitivity of video. $\hat{\textbf{y}}_{m, t}$ is a $\lvert A \rvert$-dimensional vector with the $a_{th}$ dimension fusing $\hat{m}$ and $\bm{\hat{t}}_a$, denoting the fusion of user interests and video's recency sensitivity at $A=a$.

As introduced in Section~\ref{sec:3.2}, we adopt two inference policies. Here, we provide the formulations from \textit{Policy 1} and \textit{Policy 2}:
\begin{itemize}[nolistsep, leftmargin=*]
  \item[$\bullet$] \textit{Inference Policy 1 (LDRI)}
    \begin{equation}
      \label{eqn4.8}
      \begin{aligned}
        \hat{y} = \beta \times \hat{m} + (1-\beta) \times \sigma(\bm{\hat{t}}_a)
      \end{aligned}
    \end{equation}
  \item[$\bullet$] \textit{Inference Policy 2 (LDRI-iter)}
    \begin{equation}
      \label{eqn4.9}
      \begin{split}
        & \hat{y} 
        = \sigma \{ \sum_a \left[ \beta \times \hat{m} + (1-\beta) \times \bm{\hat{t}}_a \right] \widehat{P(A=a)} \} \\
        & \widehat{P(A=a)} = \frac{\lvert \{\mathcal{D}|A=a \}\rvert }{\lvert \mathcal{D} \rvert}
      \end{split}
    \end{equation}
\end{itemize}
For both policies, $\hat{y}$ is the estimation of $P(IN|do(v))$ in Equation~(\ref{eqn3.7}), which is the implementation of backdoor adjustment, as elaborated in Sec~\ref{sec:3.2}. Hence, $\hat{y}$ delivers recommendation based on videos' causal effect on user interests.

%% file: contents/05experiments.tex
\begin{table*}
  \centering
  \caption{Overall TopK recommendation results on Pure (KuaiRand-Pure) and 1K (KuaiRand-1K). We compare different methods' performances using standard TopK recommendation metrics: RECALL@$K$, MAP@$K$, NDCG@$K$, and HR@$K$. 
  The best results are in boldface and the second best results are underlined; ** indicates $p<0.05$ compared to backbone. The Time column displays the relative running time spent on inference compared to NFM, which serves as the reference method.
  }
    
\label{tab:topk}
  \begin{center}
    {\renewcommand{\arraystretch}{0.6}
    \begin{tabular}{l l rr rr rr rr r}
      \toprule
         \multicolumn{1}{c}{\multirow{2}{*}{Dataset}} 
          & \multicolumn{1}{c}{\multirow{2}{*}{Method}} 
          & \multicolumn{2}{c}{RECALL} 
          & \multicolumn{2}{c}{MAP} 
          & \multicolumn{2}{c}{NDCG} 
          & \multicolumn{2}{c}{HR}
          & \multicolumn{1}{c}{\multirow{2}{*}{Time}}\\
            & & \multicolumn{1}{c}{Top5} & \multicolumn{1}{c}{Top10}
            & \multicolumn{1}{c}{Top5} & \multicolumn{1}{c}{Top10}
            & \multicolumn{1}{c}{Top5} & \multicolumn{1}{c}{Top10}
            & \multicolumn{1}{c}{Top5} & \multicolumn{1}{c}{Top10}  \\
      \midrule
        \multirow{21}{*}{Pure} 
          & TCCM \cite{chen2023tccm} & 0.5101 & 0.7248 & 0.3159 & 0.3612 & 0.2230 & 0.2741 & 0.6613 & 0.8351 & 1.6532 \\
          \cmidrule(lr){2-11}
          & DeepFM & 0.5077 & 0.7240 & 0.3115 & 0.3580 & 0.2219 & 0.2732 & 0.6603 & 0.8345 & 1.0092\\ 
          & DeepFM+TaFR$^{\dagger}$ \cite{wang2023measuring} & 0.5089 & 0.7247 & 0.3127 & 0.3607 & 0.2229 & 0.2740 & 0.6610 & 0.8351 & 1.1544 \\
          & DeepFM+DCR-MoE~\cite{he2023addressing} & 0.5100 & 0.7250 & 0.3149 & 0.3618 & 0.2232 & 0.2740 & 0.6610 & 0.6610 & 1.4538\\
          & DeepFM+LDRI \textbf{(Ours)}& **\underline{0.5142} & **\textbf{0.7257} & **\textbf{0.3204} & **\textbf{0.3661} & **\textbf{0.2267} & **\textbf{0.2769} & **\underline{0.6650} & **\underline{0.8362} & 1.4283\\  
          & DeepFM+LDRI-iter \textbf{(Ours)}& **\textbf{0.5148} & **\underline{0.7256} & **\underline{0.3199} & **\underline{0.3654} & **\underline{0.2266} & **\underline{0.2766} & **\textbf{0.6660}& **\textbf{0.8366} &  1.4333\\
          \cmidrule(lr){2-11}
          & NFM & 0.4952 & 0.7133 & 0.3040 & 0.3501 & 0.2171 & 0.2684 & 0.6462 & 0.8251 & 1.0000 \\
          & NFM+TaFR$^{\dagger}$ \cite{wang2023measuring} & 0.4977 & 0.7168 & 0.3072 & 0.3527 & 0.2184 & 0.2698 & 0.6488 & 0.8276 & 1.0813 \\
          & NFM+DCR-MoE~\cite{he2023addressing} & 0.4988 & 0.7188 & 0.3084 & 0.3529 & 0.2191 & 0.2704 & 0.6513 & 0.8307 & 1.3248\\
          & NFM+LDRI \textbf{(Ours)}& **0.5094 & **0.7229 & **0.3151 & **0.3612 & **0.2237 & **0.2748 & **0.6594 & **0.8355 & 1.1078\\  
          & NFM+LDRI-iter \textbf{(Ours)}& **0.5097 & **0.7233 & **0.3152 & **0.3613 & **0.2238 & **0.2748 & **0.6608 & **0.8348 & 1.1046 \\
          \cmidrule(lr){2-11}
          & AFM & 0.5037 & 0.7171 & 0.3070 & 0.3553 & 0.2193 & 0.2704 & 0.6550 & 0.8292 & 1.0245 \\
          & AFM+TaFR$^{\dagger}$ \cite{wang2023measuring} & 0.5051 & 0.7199 & 0.3082 & 0.3571 & 0.2204 & 0.2718 & 0.6577 & 0.8301 & 1.0936 \\
          & AFM+DCR-MoE~\cite{he2023addressing} & 0.5079 & 0.7208 & 0.3100 & 0.3588 & 0.2210 & 0.2722 & 0.6582 & 0.8318 & 1.6427\\
          & AFM+LDRI \textbf{(Ours)}& **0.5124 & **0.7236 & **0.3192 & **0.3651 & **0.2257 & **0.2763 & **0.6628 & **0.8331 & 1.1451\\  
          & AFM+LDRI-iter \textbf{(Ours)}& **0.5106 & **0.7236 & **0.3178 & **0.3639 & **0.2248 & **0.2756 & **0.6607 & **0.8339 & 1.1462 \\
      \midrule
        \multicolumn{1}{c}{}  & \multicolumn{1}{c}{} 
        & \multicolumn{1}{c}{Top300} & \multicolumn{1}{c}{Top500} 
        & \multicolumn{1}{c}{Top300} & \multicolumn{1}{c}{Top500} 
        & \multicolumn{1}{c}{Top300} & \multicolumn{1}{c}{Top500} 
        & \multicolumn{1}{c}{Top300} & \multicolumn{1}{c}{Top500}  \\
        \cmidrule(lr){2-11}
      \multirow{21}{*}{1K} 
        & TCCM \cite{chen2023tccm} & 0.3110 & 0.4400 & 0.0291 & 0.0389 & 0.1461 & 0.1952 & 0.9312 & 0.9677 & 3.4462 \\
        \cmidrule(lr){2-11}
        & DeepFM & 0.3095 & 0.4339 & 0.0285 & 0.0367 & 0.1431 & 0.1914 & 0.9310 & 0.9670 & 1.1874\\
        & DeepFM+TaFR$^{\dagger}$ \cite{wang2023measuring} & 0.3104 & 0.4350 & 0.0293 & 0.0379 & 0.1441 & 0.1930 & 0.9318 & 0.9681 & 1.1964\\
        & DeepFM+DCR-MoE~\cite{he2023addressing} & 0.3111 & 0.4367 & 0.0301 & 0.0387 & 0.1444 & 0.1940 & 0.9320 & 0.9685 & 2.2076\\
        & DeepFM+LDRI \textbf{(Ours)} & **0.3121 & **\underline{0.4416} & **0.0323 & **0.0409 & **0.1490 & **0.1981 & **0.9331 & **\textbf{0.9712} & 1.2462\\  
        & DeepFM+LDRI-iter \textbf{(Ours)} & **\textbf{0.3151} & **\textbf{0.4424} & **0.0327 & **0.0412 & **\underline{0.1505} & **\underline{0.1990} & **\textbf{0.9361} & **\underline{0.9691} & 1.2309 \\
        \cmidrule(lr){2-11}
        & NFM & 0.3041 & 0.4297 & 0.0301 & 0.0386 & 0.1449 & 0.1938 & 0.9269 & 0.9588 & 1.0000\\ 
        & NFM+TaFR$^{\dagger}$ \cite{wang2023measuring} & 0.3052 & 0.4308 & 0.0309 & 0.0394 & 0.1455 & 0.1950 & 0.9277 & 0.9600 & 1.1590 \\
        & NFM+DCR-MoE~\cite{he2023addressing} & 0.3060 & 0.4322 & 0.0315 & 0.0400 & 0.1467 & 0.1958 & 0.9280 & 0.9609 & 2.2143\\
        & NFM+LDRI \textbf{(Ours)} & **0.3091 & **0.4389 & **\underline{0.0327} & **\underline{0.0414} & **0.1491 & **0.1985 & **0.9300 & **0.9650 & 1.2127\\  
        & NFM+LDRI-iter \textbf{(Ours)} & **\underline{0.3135} & **0.4396 & **\textbf{0.0333} & **\textbf{0.0418} & **\textbf{0.1513} & **\textbf{0.1999} & 0.9258 & **0.9619 & 1.2201\\
        \cmidrule(lr){2-11}
        & AFM & 0.3036 & 0.4345 & 0.0282 & 0.0367 & 0.1417 & 0.1913 & 0.9331 & 0.9681 & 1.2317 \\ 
        & AFM+TaFR$^{\dagger}$ \cite{wang2023measuring} & 0.3040 & 0.4351 & 0.0290 & 0.0375 & 0.1422 & 0.1921 & 0.9333 & 0.9681 & 1.2376 \\
        & AFM+DCR-MoE~\cite{he2023addressing} & 0.3045 & 0.4363 & 0.0295 & 0.0381 & 0.1430 & 0.1931 & \underline{0.9335} & 0.9685 & 2.5714\\
        & AFM+LDRI \textbf{(Ours)}& **0.3075 & **0.4369 & **0.0317 & **0.0402 & **0.1473 & **0.1962 & 0.9269 & 0.9670 & 1.2329\\  
        & AFM+LDRI-iter \textbf{(Ours)} & **0.3076 & **0.4358 & **0.0314 & **0.0399 & **0.1471 & **0.1957 & 0.9228 & 0.9650 & 1.2400\\
    \bottomrule
    \end{tabular}
    }
    \parbox{\linewidth}{
      \raggedright
     \item[$\dagger$] We do not filter the datasets according to the experimental settings of TaFR for a fair comparison. We still outperform TaFR significantly under its settings across all TopK metrics. We do not show the comparison results for briefness.
    }
    \end{center}
    
\end{table*}

\section{Experimental Settings}
\label{sec:exp}

\subsection{Datasets}
Experiments are conducted on two large-scale short-video recommendation datasets provided by Kuaishou, Inc. \cite{gao2022kuairand}:

  \paratitle{KuaiRand-Pure} randomly selects 7,582 videos released in 3 consecutive days as candidates, which were recommended to 27,284 users in the following 30 days. There are 1.3 million interactions that reflect users' unbiased preference due to the random setting. We randomly split the dataset by  6:1:3 for training, validation, and test. 
  
  \paratitle{KuaiRand-1K} records 1,000 highly active users' over 11 million interactions on 4.3 million videos in a month. The release intervals of videos range from 0 to 1,400 days. 
  We split the dataset by interaction date, the first 20 days for training, the subsequent 3 days for validation, and the remaining for test.

Both datasets record users' interactions from April 8, 2022 to May 8, 2022. They also provide features related to users and videos encompassing user-side and video-side identification, category information, and dense attributes.

\subsection{Backbones and Compared Methods}
On top of three backbones (indicated by $\diamond$), DeepFM, NFM, and AFM, we compare four methods including TaFR and DCR-MoE (indicated by $\bullet$) and two variants of our proposed LDRI (LDRI and LDRI-iter). We also compare TCCM that considers video temporality within the model (indicated by $\circ$).

\begin{itemize}[nolistsep, leftmargin=*]
  \item[$\diamond$] \textbf{DeepFM}~\cite{guo2017deepfm} integrates the advantage of factorization machines with deep neural networks.
  \item[$\diamond$] \textbf{NFM}~\cite{he2017neural} incorporates the Bi-Interaction Pooling that is specifically crafted to capture intricate relationships.
  \item[$\diamond$] \textbf{AFM}~\cite{xiao2017attentional} employs attention mechanisms into factorization machines to prioritize important features and interactions. 
  \item[$\bullet$] \textbf{TaFR}~\cite{wang2023measuring} is a model-agnostic method that learns each item's temporality through survival analysis and combines temporality information with matching score between user and item to make prediction. TaFR calculates each item's deactivation label and sets a threshold for all items, while undesired ones are removed in both training and test stage.
  \item[$\bullet$] \textbf{DCR-MoE}~\cite{he2023addressing} eliminates the duration bias in the short video recommendation model through backdoor adjustment.
  \item[$\circ$] \textbf{TCCM}~\cite{chen2023tccm} leverages the release interval as an item-side feature to capture the temporal information of each item.
  
\end{itemize}

\subsection{Evaluation Metrics and Implementation}

For evaluation, we adopt the commonly used TopK recommendation metrics: RECALL@K, MAP@K, NDCG@K, and HR@K. We report results with $K\in\{5, 10\}$ for KuaiRand-Pure and $K\in\{300, 500\}$ for KuaiRand-1K. We also report TopK results for cold start setting.
To validate LDRI's effectiveness in deconfounding bias, we calculate average prediction scores grouped by release interval. We also report the running time compared with NFM.

For implementation, hidden dimensions of our recency perceptron are $[64, |A|]$,
the dropout rate is $0.3$, and $N$ is 1 to smooth its supervision. For both datasets, we designate $\lvert A \rvert$ as 30 days. 
We set $\alpha=0.6$ that balances the losses of two tasks, and $\beta=0.5$ which balances the inference fusion.
The above hyper-parameters are set uniformly when applied to all backbones.
Besides, we set batch size as $1024$, learning rate as $1e-4$, and optimizer as Adam \cite{Diederik2015adam}. Our code is available online: \url{https://github.com/ECNU-Text-Computing/LDRI}.

%% file: contents/06results.tex
\section{Results}
\label{sec:res}

\subsection{Overall Performance}
Table~\ref{tab:topk} presents the comprehensive comparison results based on TopK metrics and the running time spent on inference. 
In general, our proposed LDRI, a model-agnostic causal framework, outperforms all three backbones and the three state-of-the-art methods across almost all metrics. These surpassing performances underscore the significance of deconfounding release interval bias and modeling recency sensitivity. Furthermore, we have maintained a relatively low time cost. 
   
LDRI outperforms backbones as the latter learn from data collected from existing platforms and only focus on matching user and video features. Hence, they are susceptible to feedback loop bias. LDRI enhances backbones' ability to perceive recency sensitivity while matching user and item. In comparison to backbones, performance gain can also be attributed to backdoor adjustment, which eliminates confounding effect and successfully achieves unbiased recommendations.
  
TCCM fails to account for variations in recency sensitivity among different videos at the same release interval. Incorporating release interval may lead the model to capture the shortcut between short release interval and user feedback, exacerbating the confounding effect, which results in inferior performance compared with LDRI. While DCR-MoE shows promising performance, its time cost is rather high.
 
The subpar performance of TaFR is owing to its utilization of the Cox model from survival analysis to learn video temporality. This approach is inherently limited to the analysis of \textit{deactivated} videos, yet videos lack a clear label indicating whether they have deactivated or not. To address this, TaFR manually sets a uniform threshold to estimate video deactivation, which makes the model highly susceptible to threshold settings. Moreover, survival analysis solely utilizes user feedback to learn temporality, disregarding the valuable information embedded in the intrinsic features of videos. This hampers the modeling accuracy, particularly in situations where user feedback is sparse. Further, both  training and prediction stages of TaFR suffer from substantial data filtering. This renders the model impractical for real-world recommendation scenarios.

Across all datasets and metrics, both LDRI and LDRI-iter exhibit outstanding performance. Between the two, in most cases, LDRI achieves slightly better results than LDRI-iter, though the difference is not pronounced, indicating the viability of both policies.
 
\subsection{Performance across Release Intervals}
\begin{figure}
  \centering
  \subfigure[DeepFM]{
    \centering
    \includegraphics[width=0.315\columnwidth]{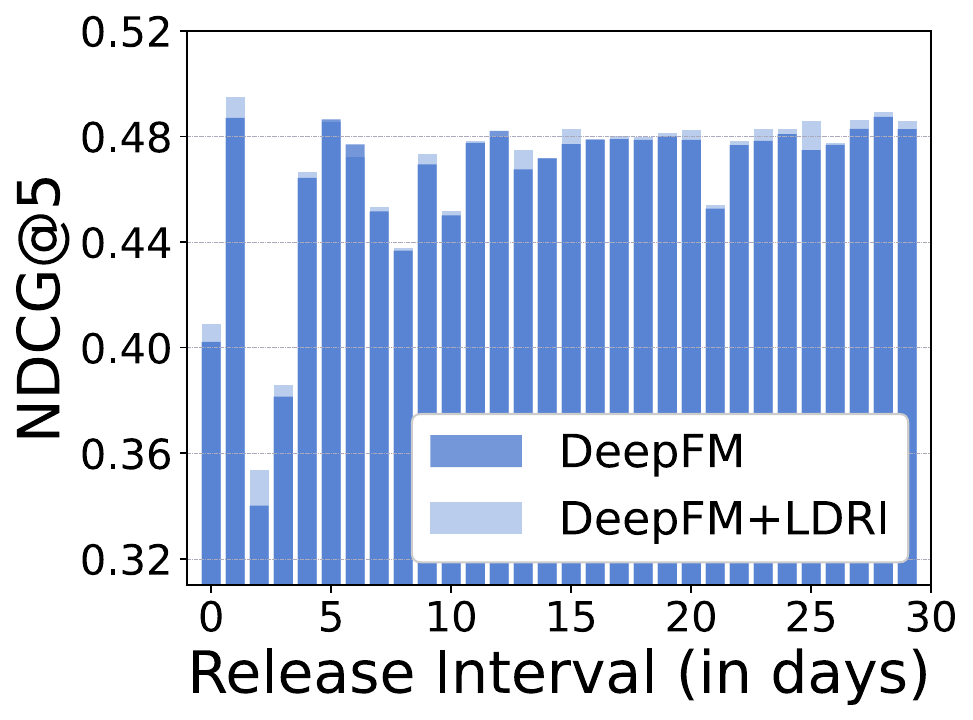}}
  \subfigure[NFM]{
    \centering
    \includegraphics[width=0.315\columnwidth]{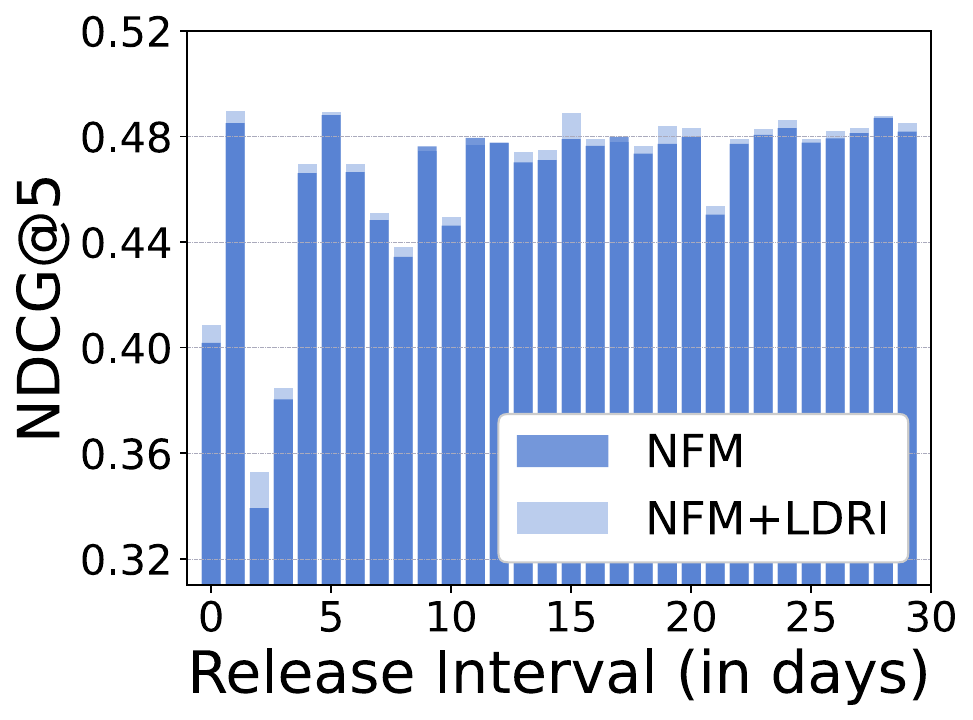}}
  \subfigure[AFM]{
    \centering
    \includegraphics[width=0.315\columnwidth]{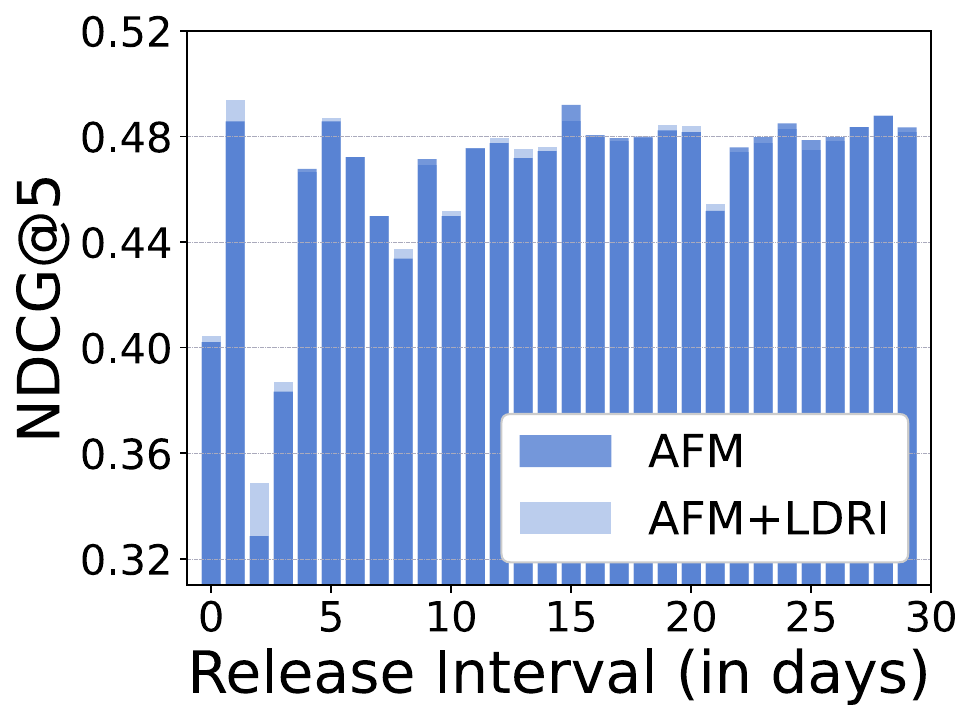}}
  \caption{NDCG@5 results at each release interval on KuaiRand-Pure with three backbones and the enhanced version with LDRI. Similar observations hold with  NDCG@300 on KuaiRand-1K.} 
    \label{fig:ndcg}
\end{figure}
We conduct an assessment to compare the improvements achieved by backbones enhanced with LDRI against backbones without LDRI. Our analysis breaks down the results at the granularity of release intervals.
Figure~\ref{fig:ndcg} provides a comparison of three backbones and their LDRI-enhanced counterparts according to NDCG@5. The application of LDRI to backbones significantly enhances their NDCG@5 across almost all release intervals, particularly notable in the case of DeepFM and AFM. The improvement can be credited to our rencency sensitivity perceptron and backdoor adjustment. LDRI precisely figures out each video's sensitivity to recency at each release interval and blocks shortcut between video and user interests, thereby augmenting recommendation performances throughout these release intervals.
It is noteworthy that the results across release intervals of backbones exhibit significant fluctuations. However, LDRI mitigates this variability, successfully enhancing both high and low values, undeniably contributes to the overall improvement of NDCG@5.

\subsection{Performance on Cold Start Setting}
\label{sec:6.3}
\begin{figure}
  \centering
  \includegraphics[width=\columnwidth]{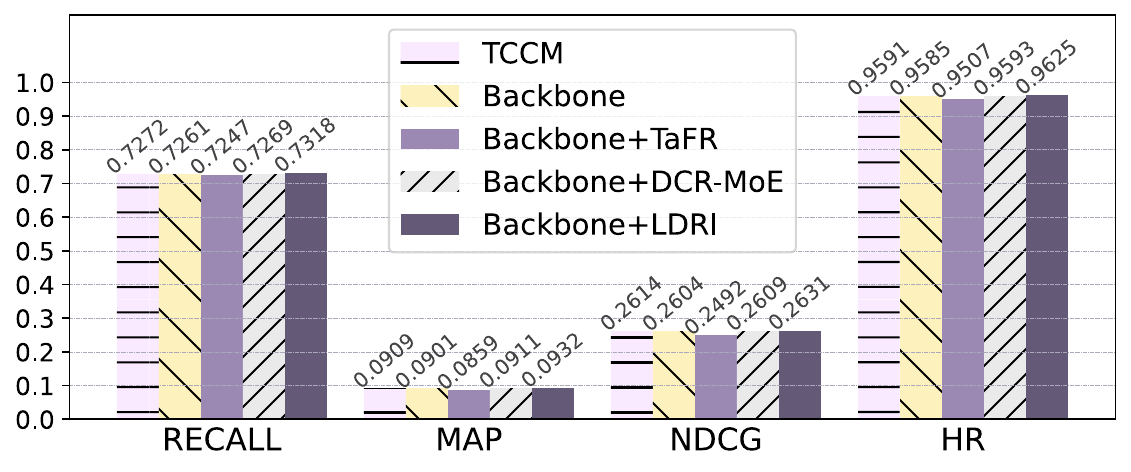}
  \caption{Metric@300 on newly released videos from KuaiRand-1K. The right four bars represent the average results of three backbones, backbones enhanced with TaFR, DCR-MoR, LDRI, respectively. Compared to backbones, the improvements by LDRI on all four metrics are significant with $p<0.05$.}
    \label{fig:cold}
\end{figure}
We evaluate the performance of three backbones, three state-of-the-art methods, along with our proposed LDRI in addressing the cold start problem.
The experiments are implemented on newly released videos that did not appear in training set. Besides, they only have interaction logs less than or equal to 2 days after release in the KuaiRand-1K dataset. The cold start experiment is not suitable for implementation on KuaiRand-Pure due to its fixed candidate pool setting. As shown in Figure~\ref{fig:cold}, LDRI significantly elevates the performance of backbones, surpassing the state-of-the-art models TCCM and DCR-MoE; however, the performance of TaFR is less satisfactory. 

In the cold start scenario, the limited feedback for newly released videos may pose challenge in learning recency sensitivity of videos. Nonetheless, this presents no impediment for LDRI, where such limitations are effectively addressed by the incorporation of video inherent features in training recency sensitivity representation, as introduced in Section~\ref{sec:4.2.2}. During the training stage, our proposed recency sensitivity perceptron profusely captures the mapping between the features of videos and their distinctive recency sensitivity representation. 
Therefore, for a newly released video $v^0$ in test set, LDRI succeeds in obtaining its accurate recency sensitivity representation $\bm{t^0}$ from prior training on analogous and early-released videos. This facilitates the derivation of all dimensions of $\bm{t^0}$. 
In contrast, TaFR enhanced backbones can only leverage existing user feedback to learn video recency sensitivity, thus its performance takes a significant hit in cold start scenario where user feedback is intensely sparse. 

\subsection{Case Study on Prediction Scores}
\begin{figure}
  \centering
  \subfigure[KuaiRand-Pure]{
    \centering
    \includegraphics[width=0.47\columnwidth]{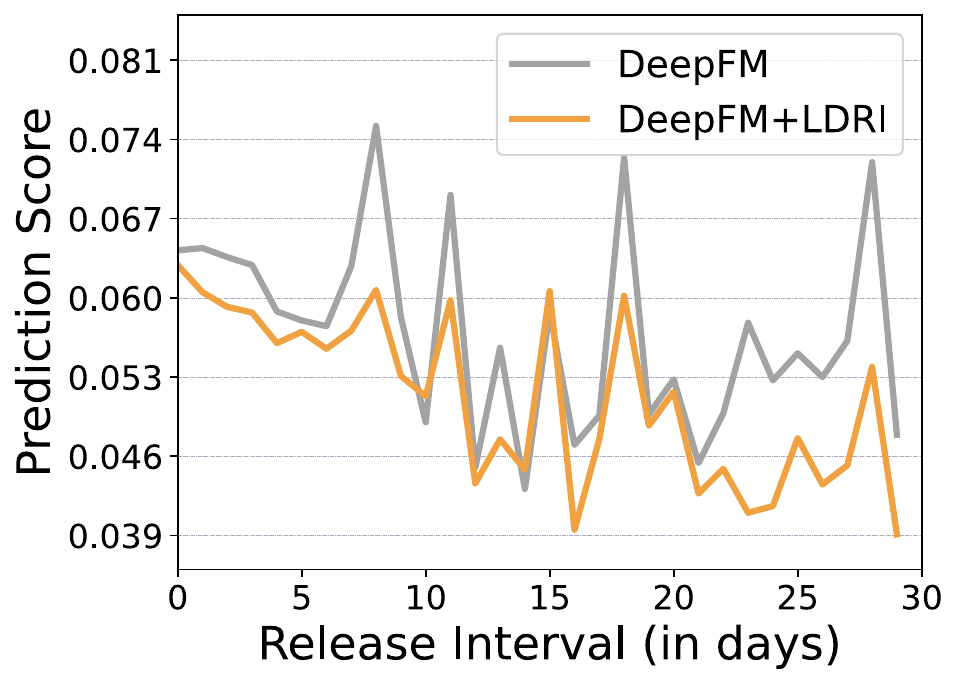}
    \label{subfig:eph_pure}}
  \subfigure[KuaiRand-1K]{
    \includegraphics[width=0.47\columnwidth]{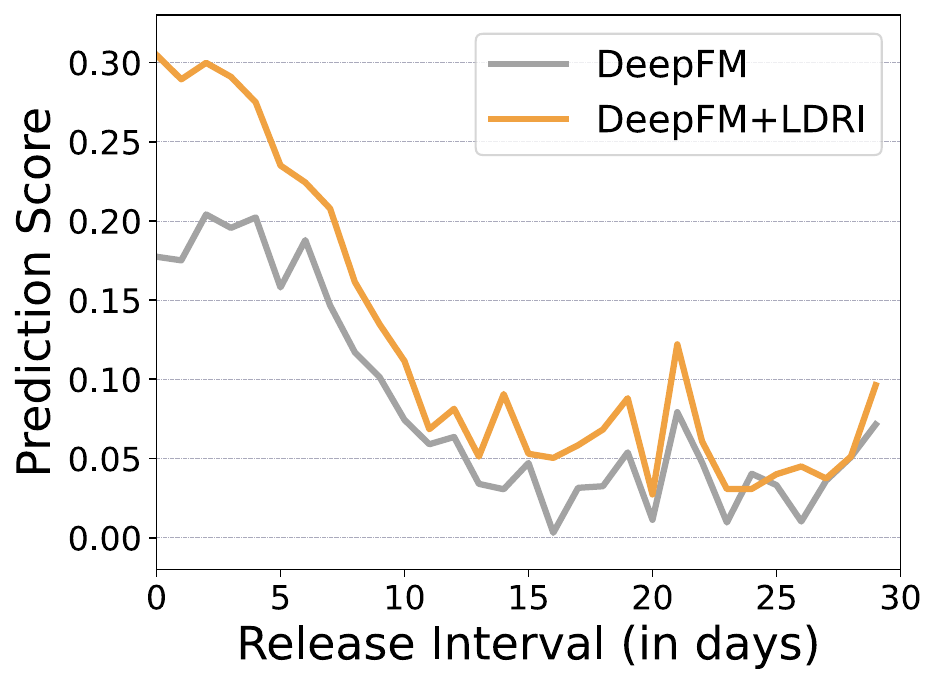}
    \label{subfig:eph_1k}}
  \caption{Average prediction scores for \textcolor{brown}{recency-sensitive} videos at each release interval. Compared to backbone, we significantly increase the disparity in prediction scores between newly released videos and previously released ones, thereby providing greater exposure opportunities for newly released videos.}
    \label{fig:pred_eph}
\end{figure}

\begin{figure}[t]
  \centering
  \subfigure[KuaiRand-Pure]{
    \centering
    \includegraphics[width=0.47\columnwidth]{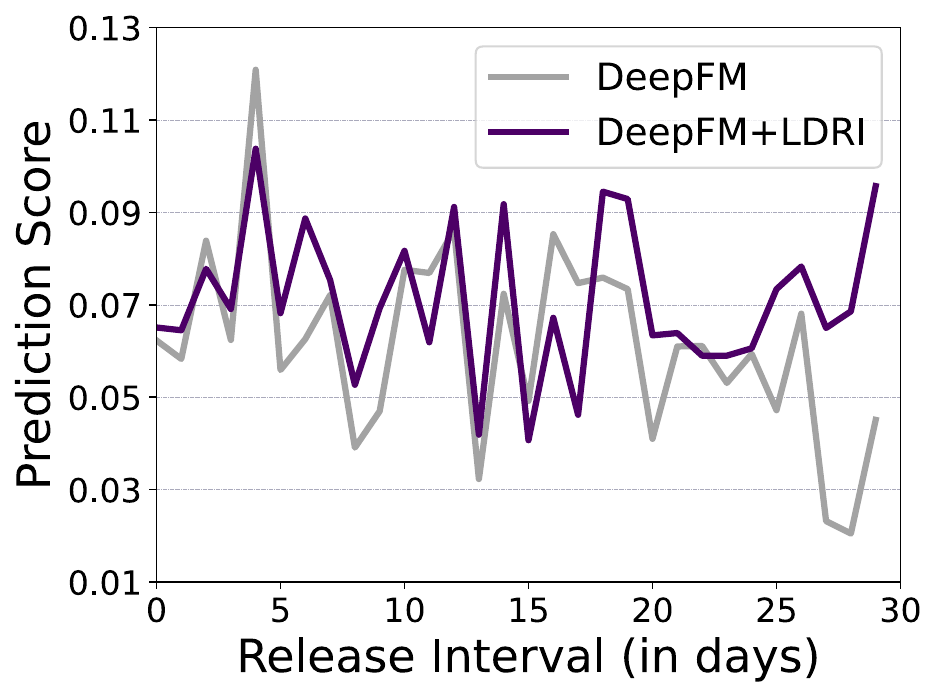}
    \label{subfig:sus_pure}}
  \subfigure[KuaiRand-1K]{
    \includegraphics[width=0.47\columnwidth]{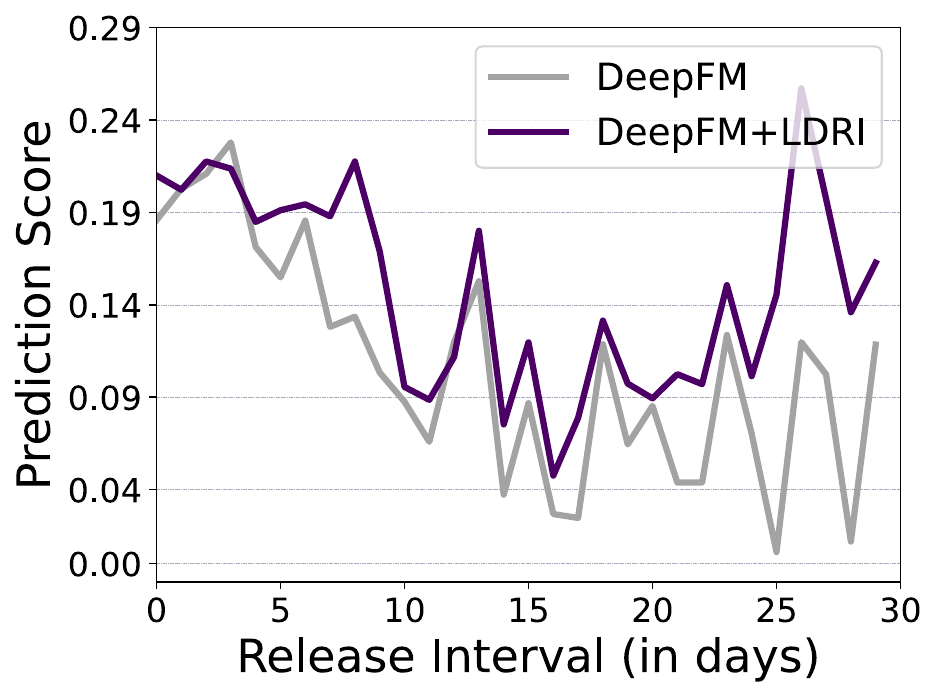}
    \label{subfig:sus_1k}}
  \caption{Average prediction scores for \textcolor{violet}{recency-insensitive} videos at each release interval. In comparison to backbone, we assign comparable prediction scores to newly released videos and those released long ago, providing them with fair opportunities for recommendation.}
    \label{fig:pred_sus}
\end{figure}

We analyze prediction scores on videos with distinct recency sensitivity obtained on  DeepFM and DeepFM enhanced with LDRI respectively, to demonstrate deconfounding capability of LDRI.
Specifically, we locate recency-sensitive videos and recency-insensitive videos from both datasets. 
Figures~\ref{fig:pred_eph} and~\ref{fig:pred_sus} respectively report the average prediction scores on recency-sensitive videos and recency-insensitive videos.

For recency-sensitive videos, given the rapid turnover and replacement of these videos, it is advisable to recommend the newest videos to users and diminish the visibility of older ones.
Thus, we anticipate higher prediction scores at shorter release intervals compared with those at longer release intervals. As depicted in Figure~\ref{subfig:eph_pure}, the prediction scores outputted by DeepFM on KuaiRand-Pure fluctuate across release intervals, even exhibiting higher prediction scores for longer release intervals.
This leads to older videos having a higher chance of exposure than newly released ones, which contradicts their recency-sensitive characteristics.
However, when applying LDRI to the backbone, there is a noticeable decrease in prediction scores for older videos. This capability serves to reduce the promotion of older videos, thereby granting more exposure to newly released ones. For results on KuaiRand-1K as illustrated in Figure~\ref{subfig:eph_1k}, while the results outputted by DeepFM show higher prediction scores for shorter release intervals compared with longer release intervals, the distinction is not pronounced. Still, when enhanced by LDRI, the distinctiveness significantly increases, which aligns well with their recency sensitivity.
  
As for recency-insensitive videos, considering their endurance over time and the slower rate of becoming outdated, we expect the prediction scores to be flat across release intervals.
As illustrated in Figure~\ref{subfig:sus_pure}, the results outputted by DeepFM show a sharp decline in prediction scores for longer release intervals. This may result in some non-outdated videos not being recommended, leading to users potentially missing out on content they find interesting and non-outdated. Consequently, creators of such thematic videos may be compelled to continuously produce new content, creating a situation of mutual loss for both users and video creators. When DeepFM is enhanced by LDRI, as confirmed in Figure~\ref{subfig:sus_1k} as well, there is a increase in scores for older videos, providing them with more opportunities for exposure, which benefits both users and video creators and corresponds to their recency-insensitive attribute.

The above results strongly demonstrate the deconfounding capability of LDRI that skillfully eradicates the release interval bias. It is crucial to emphasize that LDRI's deconfounding does not uniformly assign higher prediction scores to all older videos. Instead, it tailors recommendations based on the unique recency sensitivity of each video. 
The superior results are attributed to the robust capability of our recency sensitivity perceptron in learning the recency sensitivity of each video, coupled with the powerful deconfounding ability of the backdoor adjustment.

%% file: contents/07conclusion.tex
\section{Conclusion and Future Work}
\label{sec:conclution}

Matching models between users and videos trained on data collected from existing platforms can be susceptible to feedback loop bias. This work identifies the release interval as a confounding factor that confounds the recommender systems to prioritize newly released videos. To mitigate the confounding effect, we propose a novel model-agnostic causal architecture called LDRI. This approach jointly learns user and video matching as well as video recency sensitivity. During inference, it employs backdoor adjustments. Extensive experiments demonstrate that LDRI significantly improves the performance of backbone systems, effectively deconfounding the existing \textit{release interval} bias and providing personalized recommendations based on recency sensitivity of videos. 

One limitation of our work is that we model the dynamic changes in recency sensitivity after a video's publication, without adequately addressing the dynamic modeling of user interests. Future research should incorporate the evolving user interests over time.

%% file: recsys2024.bbl

\begin{thebibliography}{38}


\ifx \showCODEN    \undefined \def \showCODEN     #1{\unskip}     \fi
\ifx \showDOI      \undefined \def \showDOI       #1{#1}\fi
\ifx \showISBNx    \undefined \def \showISBNx     #1{\unskip}     \fi
\ifx \showISBNxiii \undefined \def \showISBNxiii  #1{\unskip}     \fi
\ifx \showISSN     \undefined \def \showISSN      #1{\unskip}     \fi
\ifx \showLCCN     \undefined \def \showLCCN      #1{\unskip}     \fi
\ifx \shownote     \undefined \def \shownote      #1{#1}          \fi
\ifx \showarticletitle \undefined \def \showarticletitle #1{#1}   \fi
\ifx \showURL      \undefined \def \showURL       {\relax}        \fi
\providecommand\bibfield[2]{#2}
\providecommand\bibinfo[2]{#2}
\providecommand\natexlab[1]{#1}
\providecommand\showeprint[2][]{arXiv:#2}

\bibitem[Cai et~al\mbox{.}(2023)]%
        {cai2023reinforcing}
\bibfield{author}{\bibinfo{person}{Qingpeng Cai}, \bibinfo{person}{Shuchang Liu}, \bibinfo{person}{Xueliang Wang}, \bibinfo{person}{Tianyou Zuo}, \bibinfo{person}{Wentao Xie}, \bibinfo{person}{Bin Yang}, \bibinfo{person}{Dong Zheng}, \bibinfo{person}{Peng Jiang}, {and} \bibinfo{person}{Kun Gai}.} \bibinfo{year}{2023}\natexlab{}.
\newblock \showarticletitle{Reinforcing User Retention in a Billion Scale Short Video Recommender System}. In \bibinfo{booktitle}{\emph{Companion Proceedings of the ACM Web Conference 2023}}. \bibinfo{pages}{421--426}.
\newblock


\bibitem[Chaney et~al\mbox{.}(2018)]%
        {chaney2018algorithmic}
\bibfield{author}{\bibinfo{person}{Allison~JB Chaney}, \bibinfo{person}{Brandon~M Stewart}, {and} \bibinfo{person}{Barbara~E Engelhardt}.} \bibinfo{year}{2018}\natexlab{}.
\newblock \showarticletitle{How algorithmic confounding in recommendation systems increases homogeneity and decreases utility}. In \bibinfo{booktitle}{\emph{Proceedings of the 12th ACM conference on recommender systems}}. \bibinfo{pages}{224--232}.
\newblock


\bibitem[Chen et~al\mbox{.}(2023a)]%
        {chen2023bias}
\bibfield{author}{\bibinfo{person}{Jiawei Chen}, \bibinfo{person}{Hande Dong}, \bibinfo{person}{Xiang Wang}, \bibinfo{person}{Fuli Feng}, \bibinfo{person}{Meng Wang}, {and} \bibinfo{person}{Xiangnan He}.} \bibinfo{year}{2023}\natexlab{a}.
\newblock \showarticletitle{Bias and debias in recommender system: A survey and future directions}.
\newblock \bibinfo{journal}{\emph{ACM Transactions on Information Systems}} \bibinfo{volume}{41}, \bibinfo{number}{3} (\bibinfo{year}{2023}), \bibinfo{pages}{1--39}.
\newblock


\bibitem[Chen et~al\mbox{.}(2023b)]%
        {chen2023tccm}
\bibfield{author}{\bibinfo{person}{Yewang Chen}, \bibinfo{person}{Weiyao Ye}, \bibinfo{person}{Guipeng Xv}, \bibinfo{person}{Chen Lin}, {and} \bibinfo{person}{Xiaomin Zhu}.} \bibinfo{year}{2023}\natexlab{b}.
\newblock \showarticletitle{TCCM: Time and Content-Aware Causal Model for Unbiased News Recommendation}. In \bibinfo{booktitle}{\emph{Proceedings of the 32nd ACM International Conference on Information and Knowledge Management}}. \bibinfo{pages}{3778--3782}.
\newblock


\bibitem[Covington et~al\mbox{.}(2016)]%
        {covington2016deep}
\bibfield{author}{\bibinfo{person}{Paul Covington}, \bibinfo{person}{Jay Adams}, {and} \bibinfo{person}{Emre Sargin}.} \bibinfo{year}{2016}\natexlab{}.
\newblock \showarticletitle{Deep neural networks for youtube recommendations}. In \bibinfo{booktitle}{\emph{Proceedings of the 10th ACM conference on recommender systems}}. \bibinfo{pages}{191--198}.
\newblock


\bibitem[Ding et~al\mbox{.}(2019)]%
        {ding2019reinforced}
\bibfield{author}{\bibinfo{person}{Jingtao Ding}, \bibinfo{person}{Yuhan Quan}, \bibinfo{person}{Xiangnan He}, \bibinfo{person}{Yong Li}, {and} \bibinfo{person}{Depeng Jin}.} \bibinfo{year}{2019}\natexlab{}.
\newblock \showarticletitle{Reinforced Negative Sampling for Recommendation with Exposure Data}. In \bibinfo{booktitle}{\emph{Proceedings of the Twenty-Eighth International Joint Conference on Artificial Intelligence, {IJCAI-19}}}. \bibinfo{publisher}{International Joint Conferences on Artificial Intelligence Organization}, \bibinfo{pages}{2230--2236}.
\newblock
\urldef\tempurl%
\url{https://doi.org/10.24963/ijcai.2019/309}
\showDOI{\tempurl}


\bibitem[Gao et~al\mbox{.}(2023)]%
        {gao2023alleviating}
\bibfield{author}{\bibinfo{person}{Chongming Gao}, \bibinfo{person}{Kexin Huang}, \bibinfo{person}{Jiawei Chen}, \bibinfo{person}{Yuan Zhang}, \bibinfo{person}{Biao Li}, \bibinfo{person}{Peng Jiang}, \bibinfo{person}{Shiqin Wang}, \bibinfo{person}{Zhong Zhang}, {and} \bibinfo{person}{Xiangnan He}.} \bibinfo{year}{2023}\natexlab{}.
\newblock \showarticletitle{Alleviating Matthew Effect of Offline Reinforcement Learning in Interactive Recommendation}.
\newblock \bibinfo{journal}{\emph{Proceedings of the 46th International ACM SIGIR Conference on Research and Development in Information Retrieval}} (\bibinfo{year}{2023}).
\newblock
\urldef\tempurl%
\url{https://api.semanticscholar.org/CorpusID:259501995}
\showURL{%
\tempurl}


\bibitem[Gao et~al\mbox{.}(2022)]%
        {gao2022kuairand}
\bibfield{author}{\bibinfo{person}{Chongming Gao}, \bibinfo{person}{Shijun Li}, \bibinfo{person}{Yuan Zhang}, \bibinfo{person}{Jiawei Chen}, \bibinfo{person}{Biao Li}, \bibinfo{person}{Wenqiang Lei}, \bibinfo{person}{Peng Jiang}, {and} \bibinfo{person}{Xiangnan He}.} \bibinfo{year}{2022}\natexlab{}.
\newblock \showarticletitle{KuaiRand: An Unbiased Sequential Recommendation Dataset with Randomly Exposed Videos}. In \bibinfo{booktitle}{\emph{Proceedings of the 31st ACM International Conference on Information \& Knowledge Management}}. \bibinfo{pages}{3953--3957}.
\newblock


\bibitem[Guo et~al\mbox{.}(2017)]%
        {guo2017deepfm}
\bibfield{author}{\bibinfo{person}{Huifeng Guo}, \bibinfo{person}{Ruiming TANG}, \bibinfo{person}{Yunming Ye}, \bibinfo{person}{Zhenguo Li}, {and} \bibinfo{person}{Xiuqiang He}.} \bibinfo{year}{2017}\natexlab{}.
\newblock \showarticletitle{DeepFM: A Factorization-Machine based Neural Network for CTR Prediction}. In \bibinfo{booktitle}{\emph{Proceedings of the Twenty-Sixth International Joint Conference on Artificial Intelligence, {IJCAI-17}}}. \bibinfo{pages}{1725--1731}.
\newblock
\urldef\tempurl%
\url{https://doi.org/10.24963/ijcai.2017/239}
\showDOI{\tempurl}


\bibitem[He et~al\mbox{.}(2016)]%
        {he2016deep}
\bibfield{author}{\bibinfo{person}{Kaiming He}, \bibinfo{person}{Xiangyu Zhang}, \bibinfo{person}{Shaoqing Ren}, {and} \bibinfo{person}{Jian Sun}.} \bibinfo{year}{2016}\natexlab{}.
\newblock \showarticletitle{Deep residual learning for image recognition}. In \bibinfo{booktitle}{\emph{Proceedings of the IEEE conference on computer vision and pattern recognition}}. \bibinfo{pages}{770--778}.
\newblock


\bibitem[He et~al\mbox{.}(2022)]%
        {he2022causal}
\bibfield{author}{\bibinfo{person}{Ming He}, \bibinfo{person}{Xin Chen}, \bibinfo{person}{Xinlei Hu}, {and} \bibinfo{person}{Changshu Li}.} \bibinfo{year}{2022}\natexlab{}.
\newblock \showarticletitle{Causal intervention for sentiment de-biasing in recommendation}. In \bibinfo{booktitle}{\emph{Proceedings of the 31st ACM International Conference on Information \& Knowledge Management}}. \bibinfo{pages}{4014--4018}.
\newblock


\bibitem[He and Chua(2017)]%
        {he2017neural}
\bibfield{author}{\bibinfo{person}{Xiangnan He} {and} \bibinfo{person}{Tat-Seng Chua}.} \bibinfo{year}{2017}\natexlab{}.
\newblock \showarticletitle{Neural factorization machines for sparse predictive analytics}. In \bibinfo{booktitle}{\emph{Proceedings of the 40th International ACM SIGIR conference on Research and Development in Information Retrieval}}. \bibinfo{pages}{355--364}.
\newblock


\bibitem[He et~al\mbox{.}(2023)]%
        {he2023addressing}
\bibfield{author}{\bibinfo{person}{Xiangnan He}, \bibinfo{person}{Yang Zhang}, \bibinfo{person}{Fuli Feng}, \bibinfo{person}{Chonggang Song}, \bibinfo{person}{Lingling Yi}, \bibinfo{person}{Guohui Ling}, {and} \bibinfo{person}{Yongdong Zhang}.} \bibinfo{year}{2023}\natexlab{}.
\newblock \showarticletitle{Addressing confounding feature issue for causal recommendation}.
\newblock \bibinfo{journal}{\emph{ACM Transactions on Information Systems}} \bibinfo{volume}{41}, \bibinfo{number}{3} (\bibinfo{year}{2023}), \bibinfo{pages}{1--23}.
\newblock


\bibitem[Huang et~al\mbox{.}(2022)]%
        {10.1145/3488560.3498375}
\bibfield{author}{\bibinfo{person}{Jin Huang}, \bibinfo{person}{Harrie Oosterhuis}, {and} \bibinfo{person}{Maarten de Rijke}.} \bibinfo{year}{2022}\natexlab{}.
\newblock \showarticletitle{It Is Different When Items Are Older: Debiasing Recommendations When Selection Bias and User Preferences Are Dynamic}. In \bibinfo{booktitle}{\emph{Proceedings of the Fifteenth ACM International Conference on Web Search and Data Mining}} (Virtual Event, AZ, USA) \emph{(\bibinfo{series}{WSDM '22})}. \bibinfo{publisher}{Association for Computing Machinery}, \bibinfo{address}{New York, NY, USA}, \bibinfo{pages}{381–389}.
\newblock
\showISBNx{9781450391320}
\urldef\tempurl%
\url{https://doi.org/10.1145/3488560.3498375}
\showDOI{\tempurl}


\bibitem[Huang et~al\mbox{.}(2023)]%
        {huang2023personal}
\bibfield{author}{\bibinfo{person}{Zhenya Huang}, \bibinfo{person}{Binbin Jin}, \bibinfo{person}{Hongke Zhao}, \bibinfo{person}{Qi Liu}, \bibinfo{person}{Defu Lian}, \bibinfo{person}{Bao Tengfei}, {and} \bibinfo{person}{Enhong Chen}.} \bibinfo{year}{2023}\natexlab{}.
\newblock \showarticletitle{Personal or general? a hybrid strategy with multi-factors for news recommendation}.
\newblock \bibinfo{journal}{\emph{ACM Transactions on Information Systems}} \bibinfo{volume}{41}, \bibinfo{number}{2} (\bibinfo{year}{2023}), \bibinfo{pages}{1--29}.
\newblock


\bibitem[Jenkins(2005)]%
        {jenkins2005survival}
\bibfield{author}{\bibinfo{person}{Stephen~P Jenkins}.} \bibinfo{year}{2005}\natexlab{}.
\newblock \showarticletitle{Survival analysis}.
\newblock \bibinfo{journal}{\emph{Unpublished manuscript, Institute for Social and Economic Research, University of Essex, Colchester, UK}}  \bibinfo{volume}{42} (\bibinfo{year}{2005}), \bibinfo{pages}{54--56}.
\newblock


\bibitem[Jeunen(2023)]%
        {jeunen2023probabilistic}
\bibfield{author}{\bibinfo{person}{Olivier Jeunen}.} \bibinfo{year}{2023}\natexlab{}.
\newblock \showarticletitle{A Probabilistic Position Bias Model for Short-Video Recommendation Feeds}. In \bibinfo{booktitle}{\emph{Proceedings of the 17th ACM Conference on Recommender Systems}}. \bibinfo{pages}{675--681}.
\newblock


\bibitem[Kingma and Ba(2015)]%
        {Diederik2015adam}
\bibfield{author}{\bibinfo{person}{Diederik~P. Kingma} {and} \bibinfo{person}{Jimmy Ba}.} \bibinfo{year}{2015}\natexlab{}.
\newblock \showarticletitle{Adam: {A} Method for Stochastic Optimization}. In \bibinfo{booktitle}{\emph{3rd International Conference on Learning Representations, {ICLR} 2015, San Diego, CA, USA, May 7-9, 2015, Conference Track Proceedings}}, \bibfield{editor}{\bibinfo{person}{Yoshua Bengio} {and} \bibinfo{person}{Yann LeCun}} (Eds.).
\newblock
\urldef\tempurl%
\url{http://arxiv.org/abs/1412.6980}
\showURL{%
\tempurl}


\bibitem[Lin et~al\mbox{.}(2021)]%
        {lin2021mitigating}
\bibfield{author}{\bibinfo{person}{Chen Lin}, \bibinfo{person}{Xinyi Liu}, \bibinfo{person}{Guipeng Xv}, {and} \bibinfo{person}{Hui Li}.} \bibinfo{year}{2021}\natexlab{}.
\newblock \showarticletitle{Mitigating sentiment bias for recommender systems}. In \bibinfo{booktitle}{\emph{Proceedings of the 44th International ACM SIGIR Conference on Research and Development in Information Retrieval}}. \bibinfo{pages}{31--40}.
\newblock


\bibitem[Lin et~al\mbox{.}(2023)]%
        {lin2023tree}
\bibfield{author}{\bibinfo{person}{Xiao Lin}, \bibinfo{person}{Xiaokai Chen}, \bibinfo{person}{Linfeng Song}, \bibinfo{person}{Jingwei Liu}, \bibinfo{person}{Biao Li}, {and} \bibinfo{person}{Peng Jiang}.} \bibinfo{year}{2023}\natexlab{}.
\newblock \showarticletitle{Tree based Progressive Regression Model for Watch-Time Prediction in Short-video Recommendation}. In \bibinfo{booktitle}{\emph{Proceedings of the 29th ACM SIGKDD Conference on Knowledge Discovery and Data Mining}} \emph{(\bibinfo{series}{KDD '23})}. \bibinfo{publisher}{Association for Computing Machinery}, \bibinfo{pages}{4497–4506}.
\newblock
\showISBNx{9798400701030}


\bibitem[Lin et~al\mbox{.}(2022)]%
        {lin2022feature}
\bibfield{author}{\bibinfo{person}{Zihan Lin}, \bibinfo{person}{Hui Wang}, \bibinfo{person}{Jingshu Mao}, \bibinfo{person}{Wayne~Xin Zhao}, \bibinfo{person}{Cheng Wang}, \bibinfo{person}{Peng Jiang}, {and} \bibinfo{person}{Ji-Rong Wen}.} \bibinfo{year}{2022}\natexlab{}.
\newblock \showarticletitle{Feature-aware diversified re-ranking with disentangled representations for relevant recommendation}. In \bibinfo{booktitle}{\emph{Proceedings of the 28th ACM SIGKDD Conference on Knowledge Discovery and Data Mining}}. \bibinfo{pages}{3327--3335}.
\newblock


\bibitem[Mansoury et~al\mbox{.}(2020)]%
        {mansoury2020feedback}
\bibfield{author}{\bibinfo{person}{Masoud Mansoury}, \bibinfo{person}{Himan Abdollahpouri}, \bibinfo{person}{Mykola Pechenizkiy}, \bibinfo{person}{Bamshad Mobasher}, {and} \bibinfo{person}{Robin Burke}.} \bibinfo{year}{2020}\natexlab{}.
\newblock \showarticletitle{Feedback loop and bias amplification in recommender systems}. In \bibinfo{booktitle}{\emph{Proceedings of the 29th ACM international conference on information \& knowledge management}}. \bibinfo{pages}{2145--2148}.
\newblock


\bibitem[Pearl(2009)]%
        {pearl2009causality}
\bibfield{author}{\bibinfo{person}{Judea Pearl}.} \bibinfo{year}{2009}\natexlab{}.
\newblock \bibinfo{booktitle}{\emph{Causality}}.
\newblock \bibinfo{publisher}{Cambridge university press}.
\newblock


\bibitem[Quan et~al\mbox{.}(2023)]%
        {quan2023alleviating}
\bibfield{author}{\bibinfo{person}{Yuhan Quan}, \bibinfo{person}{Jingtao Ding}, \bibinfo{person}{Chen Gao}, \bibinfo{person}{Nian Li}, \bibinfo{person}{Lingling Yi}, \bibinfo{person}{Depeng Jin}, {and} \bibinfo{person}{Yong Li}.} \bibinfo{year}{2023}\natexlab{}.
\newblock \showarticletitle{Alleviating Video-length Effect for Micro-video Recommendation}.
\newblock \bibinfo{journal}{\emph{ACM Transactions on Information Systems}} \bibinfo{volume}{42}, \bibinfo{number}{2} (\bibinfo{year}{2023}), \bibinfo{pages}{1--24}.
\newblock


\bibitem[Shi et~al\mbox{.}(2023)]%
        {shi2023relieving}
\bibfield{author}{\bibinfo{person}{Xiaoyu Shi}, \bibinfo{person}{Quanliang Liu}, \bibinfo{person}{Hong Xie}, \bibinfo{person}{Di Wu}, \bibinfo{person}{Bo Peng}, \bibinfo{person}{MingSheng Shang}, {and} \bibinfo{person}{Defu Lian}.} \bibinfo{year}{2023}\natexlab{}.
\newblock \showarticletitle{Relieving Popularity Bias in Interactive Recommendation: A Diversity-Novelty-Aware Reinforcement Learning Approach}.
\newblock \bibinfo{journal}{\emph{ACM Transactions on Information Systems}} \bibinfo{volume}{42}, \bibinfo{number}{2} (\bibinfo{year}{2023}), \bibinfo{pages}{1--30}.
\newblock


\bibitem[Shutsko(2020)]%
        {shutsko2020user}
\bibfield{author}{\bibinfo{person}{Aliaksandra Shutsko}.} \bibinfo{year}{2020}\natexlab{}.
\newblock \showarticletitle{User-generated short video content in social media. A case study of TikTok}. In \bibinfo{booktitle}{\emph{Social Computing and Social Media. Participation, User Experience, Consumer Experience, and Applications of Social Computing: 12th International Conference, SCSM 2020, Held as Part of the 22nd HCI International Conference, HCII 2020, Copenhagen, Denmark, July 19--24, 2020, Proceedings, Part II 22}}. Springer, \bibinfo{pages}{108--125}.
\newblock


\bibitem[Tang et~al\mbox{.}(2023)]%
        {tang2023counterfactual}
\bibfield{author}{\bibinfo{person}{Shisong Tang}, \bibinfo{person}{Qing Li}, \bibinfo{person}{Dingmin Wang}, \bibinfo{person}{Ci Gao}, \bibinfo{person}{Wentao Xiao}, \bibinfo{person}{Dan Zhao}, \bibinfo{person}{Yong Jiang}, \bibinfo{person}{Qian Ma}, {and} \bibinfo{person}{Aoyang Zhang}.} \bibinfo{year}{2023}\natexlab{}.
\newblock \showarticletitle{Counterfactual Video Recommendation for Duration Debiasing}. In \bibinfo{booktitle}{\emph{Proceedings of the 29th ACM SIGKDD Conference on Knowledge Discovery and Data Mining}}. \bibinfo{pages}{4894--4903}.
\newblock


\bibitem[Therneau et~al\mbox{.}(2000)]%
        {therneau2000cox}
\bibfield{author}{\bibinfo{person}{Terry~M Therneau}, \bibinfo{person}{Patricia~M Grambsch}, \bibinfo{person}{Terry~M Therneau}, {and} \bibinfo{person}{Patricia~M Grambsch}.} \bibinfo{year}{2000}\natexlab{}.
\newblock \bibinfo{booktitle}{\emph{The cox model}}.
\newblock \bibinfo{publisher}{Springer}.
\newblock


\bibitem[Tian et~al\mbox{.}(2022)]%
        {tian2022multi}
\bibfield{author}{\bibinfo{person}{Yu Tian}, \bibinfo{person}{Jianxin Chang}, \bibinfo{person}{Yanan Niu}, \bibinfo{person}{Yang Song}, {and} \bibinfo{person}{Chenliang Li}.} \bibinfo{year}{2022}\natexlab{}.
\newblock \showarticletitle{When multi-level meets multi-interest: A multi-grained neural model for sequential recommendation}. In \bibinfo{booktitle}{\emph{Proceedings of the 45th International ACM SIGIR Conference on Research and Development in Information Retrieval}}. \bibinfo{pages}{1632--1641}.
\newblock


\bibitem[Wang et~al\mbox{.}(2023)]%
        {wang2023measuring}
\bibfield{author}{\bibinfo{person}{Jiayin Wang}, \bibinfo{person}{Weizhi Ma}, \bibinfo{person}{Chumeng Jiang}, \bibinfo{person}{Min Zhang}, \bibinfo{person}{Yuan Zhang}, \bibinfo{person}{Biao Li}, {and} \bibinfo{person}{Peng Jiang}.} \bibinfo{year}{2023}\natexlab{}.
\newblock \showarticletitle{Measuring Item Global Residual Value for Fair Recommendation}. In \bibinfo{booktitle}{\emph{Proceedings of the 46th International ACM SIGIR Conference on Research and Development in Information Retrieval}}. \bibinfo{pages}{269--278}.
\newblock


\bibitem[Wang et~al\mbox{.}(2019)]%
        {wang2019machine}
\bibfield{author}{\bibinfo{person}{Ping Wang}, \bibinfo{person}{Yan Li}, {and} \bibinfo{person}{Chandan~K Reddy}.} \bibinfo{year}{2019}\natexlab{}.
\newblock \showarticletitle{Machine learning for survival analysis: A survey}.
\newblock \bibinfo{journal}{\emph{ACM Computing Surveys (CSUR)}} \bibinfo{volume}{51}, \bibinfo{number}{6} (\bibinfo{year}{2019}), \bibinfo{pages}{1--36}.
\newblock


\bibitem[Wang et~al\mbox{.}(2021)]%
        {wang2021deconfounded}
\bibfield{author}{\bibinfo{person}{Wenjie Wang}, \bibinfo{person}{Fuli Feng}, \bibinfo{person}{Xiangnan He}, \bibinfo{person}{Xiang Wang}, {and} \bibinfo{person}{Tat-Seng Chua}.} \bibinfo{year}{2021}\natexlab{}.
\newblock \showarticletitle{Deconfounded recommendation for alleviating bias amplification}. In \bibinfo{booktitle}{\emph{Proceedings of the 27th ACM SIGKDD Conference on Knowledge Discovery \& Data Mining}}. \bibinfo{pages}{1717--1725}.
\newblock


\bibitem[Wei et~al\mbox{.}(2021)]%
        {wei2021model}
\bibfield{author}{\bibinfo{person}{Tianxin Wei}, \bibinfo{person}{Fuli Feng}, \bibinfo{person}{Jiawei Chen}, \bibinfo{person}{Ziwei Wu}, \bibinfo{person}{Jinfeng Yi}, {and} \bibinfo{person}{Xiangnan He}.} \bibinfo{year}{2021}\natexlab{}.
\newblock \showarticletitle{Model-agnostic counterfactual reasoning for eliminating popularity bias in recommender system}. In \bibinfo{booktitle}{\emph{Proceedings of the 27th ACM SIGKDD Conference on Knowledge Discovery \& Data Mining}}. \bibinfo{pages}{1791--1800}.
\newblock


\bibitem[Xiao et~al\mbox{.}(2017)]%
        {xiao2017attentional}
\bibfield{author}{\bibinfo{person}{Jun Xiao}, \bibinfo{person}{Hao Ye}, \bibinfo{person}{Xiangnan He}, \bibinfo{person}{Hanwang Zhang}, \bibinfo{person}{Fei Wu}, {and} \bibinfo{person}{Tat-Seng Chua}.} \bibinfo{year}{2017}\natexlab{}.
\newblock \showarticletitle{Attentional Factorization Machines: Learning the Weight of Feature Interactions via Attention Networks}. In \bibinfo{booktitle}{\emph{Proceedings of the Twenty-Sixth International Joint Conference on Artificial Intelligence, {IJCAI-17}}}. \bibinfo{pages}{3119--3125}.
\newblock
\urldef\tempurl%
\url{https://doi.org/10.24963/ijcai.2017/435}
\showDOI{\tempurl}


\bibitem[Zhan et~al\mbox{.}(2022)]%
        {zhan2022deconfounding}
\bibfield{author}{\bibinfo{person}{Ruohan Zhan}, \bibinfo{person}{Changhua Pei}, \bibinfo{person}{Qiang Su}, \bibinfo{person}{Jianfeng Wen}, \bibinfo{person}{Xueliang Wang}, \bibinfo{person}{Guanyu Mu}, \bibinfo{person}{Dong Zheng}, \bibinfo{person}{Peng Jiang}, {and} \bibinfo{person}{Kun Gai}.} \bibinfo{year}{2022}\natexlab{}.
\newblock \showarticletitle{Deconfounding Duration Bias in Watch-time Prediction for Video Recommendation}. In \bibinfo{booktitle}{\emph{Proceedings of the 28th ACM SIGKDD Conference on Knowledge Discovery and Data Mining}}. \bibinfo{pages}{4472--4481}.
\newblock


\bibitem[Zhang et~al\mbox{.}(2017)]%
        {zhang2017building}
\bibfield{author}{\bibinfo{person}{Bo-Wen Zhang}, \bibinfo{person}{Xu-Cheng Yin}, \bibinfo{person}{Fang Zhou}, {and} \bibinfo{person}{Jian-Lin Jin}.} \bibinfo{year}{2017}\natexlab{}.
\newblock \showarticletitle{Building your own reading list anytime via embedding relevance, quality, timeliness and diversity}. In \bibinfo{booktitle}{\emph{Proceedings of the 40th International ACM SIGIR Conference on Research and Development in Information Retrieval}}. \bibinfo{pages}{1109--1112}.
\newblock


\bibitem[Zhang et~al\mbox{.}(2021)]%
        {zhang2021causal}
\bibfield{author}{\bibinfo{person}{Yang Zhang}, \bibinfo{person}{Fuli Feng}, \bibinfo{person}{Xiangnan He}, \bibinfo{person}{Tianxin Wei}, \bibinfo{person}{Chonggang Song}, \bibinfo{person}{Guohui Ling}, {and} \bibinfo{person}{Yongdong Zhang}.} \bibinfo{year}{2021}\natexlab{}.
\newblock \showarticletitle{Causal intervention for leveraging popularity bias in recommendation}. In \bibinfo{booktitle}{\emph{Proceedings of the 44th International ACM SIGIR Conference on Research and Development in Information Retrieval}}. \bibinfo{pages}{11--20}.
\newblock


\bibitem[Zheng et~al\mbox{.}(2022)]%
        {zheng2022dvr}
\bibfield{author}{\bibinfo{person}{Yu Zheng}, \bibinfo{person}{Chen Gao}, \bibinfo{person}{Jingtao Ding}, \bibinfo{person}{Lingling Yi}, \bibinfo{person}{Depeng Jin}, \bibinfo{person}{Yong Li}, {and} \bibinfo{person}{Meng Wang}.} \bibinfo{year}{2022}\natexlab{}.
\newblock \showarticletitle{Dvr: micro-video recommendation optimizing watch-time-gain under duration bias}. In \bibinfo{booktitle}{\emph{Proceedings of the 30th ACM International Conference on Multimedia}}. \bibinfo{pages}{334--345}.
\newblock


\end{thebibliography}
